\documentclass[prl,twocolumn,10pt,aps]{revtex4-2}
\usepackage[utf8]{inputenc}
\usepackage[normalem]{ulem}
\usepackage{textcomp}
\usepackage{amssymb}
\usepackage{natbib}
\usepackage{amsmath}
\usepackage{amsfonts}
\usepackage{graphicx}
\usepackage{subfigure}
\usepackage{mathrsfs}
\usepackage{dcolumn} 
\usepackage[T1]{fontenc}
\usepackage{bm}
\usepackage{color}
\usepackage{lipsum}
\usepackage{cancel}
\usepackage{mathbbol}
\usepackage{epsfig}
\usepackage{units}
\usepackage{esint}
\usepackage{soul} 
\usepackage{afterpage}
\usepackage{hyperref}
\usepackage{braket}
\newcommand{\me}[1]{\left\langle #1 \right\rangle }

\usepackage{xurl}
\usepackage{hyperref}

\begin{document}
	\title{Universality and weak-ergodicity breaking in quantum quenches} 
	\author{Guido Giachetti$^{1}$}
    \author{Andrea Solfanelli$^{2}$}
	\author{Nicol\`o Defenu$^{3,4}$}
	\affiliation{$^{1}$ Laboratoire de Physique de l'\'Ecole Normale Sup\'erieure, CNRS, ENS $\&$ PSL University, Sorbonne Universit\'e, Universit\'e Paris Cité, 75005 Paris, France}
    \affiliation{$^{2}$Max Planck Institute for the Physics of Complex Systems, Nöthnitzer Str. 38, 01187 Dresden, Germany}
	\affiliation{$^{3}$Institut f\"ur Theoretische Physik, ETH Z\"urich, Wolfgang-Pauli-Str. 27 Z\"urich, Switzerland}
    \affiliation{$^{4}$CNR-INO, Area Science Park, Basovizza, I-34149 Trieste, Italy}
	\begin{abstract}
	\noindent
Understanding equilibration and universality after sudden quantum quenches remains a central challenge in isolated many-body systems. Persistent oscillations and anomalous scaling reported in lattice models appear to challenge the standard picture based on integrability and quantum-to-classical correspondence. Focusing on the quantum $O(n)$ model in the large-$n$ limit, we show that these apparent anomalies originate from lattice effects and an underlying integrable structure. \textcolor{black}{In particular, we map the post-quench dynamics onto an integrable Neumann system and we derive a new action-angle description valid in the thermodynamic limit. This framework explains both the convergence of long-time averages to the generalized Gibbs ensemble and the emergence of persistent oscillations associated with an isolated mode at the upper edge of the phonon spectrum. We further show that the expected thermal universality class is recovered in the appropriate scaling regime, while the oscillations disappear in the quantum field theory limit.}
 
 \end{abstract}
	\maketitle 
\noindent
\emph{Introduction:}
 The approach to equilibrium in isolated quantum systems is a central problem in modern statistical mechanics~\cite{kubo1991nonequilibrium}. While thermalization in classical systems is typically attributed to dynamical chaos~\cite{schuster1988deterministic,gallavotti2007fermi}, quantum systems evolve unitarily and do not admit a direct classical notion of chaos. Nevertheless, a statistical description of equilibration has been established through the eigenstate thermalization hypothesis (ETH)~\cite{vonNeumann2010proof,srednicki1994chaos,rigol2012alternatives,goldstein2006canonical}, whose broad validity in interacting systems has been extensively investigated~\cite{manmana2007strongly,moeckel2008interaction,biroli2010effect,khatami2013fluctuation,yoshizawa2018numerical,foini2019eigenstate,pappalardi2022eigenstate,pappalardi2025full}.

Quantum quenches provide a particularly powerful protocol to probe equilibration and universality in isolated many-body systems~\cite{mitra2018quantum,polkovnikov2011colloquium}. Beyond asymptotic thermalization, quenches often display universal dynamical scaling~\cite{calabrese2006time,calabrese2007quantum} and universal energy statistics~\cite{gambassi2012large,solfanelli2025universal}, reflecting the universality class of the associated equilibrium critical point via the quantum-to-classical correspondence~\cite{gambassi2011quantum}. Integrable systems play a crucial role in this context: although they do not thermalize due to the presence of infinitely many conserved quantities, they can equilibrate through dephasing—often termed kinematical chaos~\cite{lasinio1996chaotic,lenci1996ergodic,graffi1996ergodic}—leading to relaxation towards a Generalized Gibbs Ensemble (GGE)~\cite{rigol2007relaxation,rigol2009breakdown,calabrese2011quantum,rigol2016fundamental,Ilievski2015complete,biagetti2408generalised}. In contrast, long-range interactions can violate this mechanism and produce persistent oscillations~\cite{defenu2021metastability,giachetti2021entanglement,giachetti2025conditions}.

Recent studies have reported persistent oscillations in local lattice systems after quantum quenches, interpreted as weak ergodicity breaking~\cite{banuls2011strong,delfino2017theory,rakovszky2016hamiltonian,kormos2017real,collura2018dynamical,robinson2019signatures,castro2020entanglement,scopa2022entanglement}. In parallel, scaling behavior apparently incompatible with thermal universality has been observed in the large-$n$ $O(n)$ model~\cite{sciolla2013quantum,weidinger2017dynamical}, while the applicability of the GGE in this setting has been questioned~\cite{chandran2013equilibration}. If taken at face value, these findings would suggest the emergence of unconventional dynamical universality classes, intermediate between integrability and thermalization, or genuinely non-thermal critical behavior.

In this work, we show that these seemingly anomalous features admit a coherent and unified interpretation once lattice effects and the full dynamical structure of the theory are properly accounted for. Focusing on the $O(n)$ symmetric model in the large-$n$ limit—a paradigmatic and exactly solvable setting for non-equilibrium field theory~\cite{moshe2003quantum,sotiriadi2010quantum,gambassi2011quantum}— \textcolor{black}{we demonstrate that the model is integrable and exhibits weak equilibration to a generalized Gibbs ensemble (GGE) - meaning that long-time averages converge to the GGE predictions - } resolving the puzzle raised in Ref.~\cite{chandran2013equilibration}. We further show that the expected thermal universality class~\cite{vojta1996quantum} is recovered in the appropriate dynamical scaling regime, while lattice-induced crossover effects account for the deviations reported in Refs.~\cite{sciolla2013quantum,weidinger2017dynamical}.

\textcolor{black}{ Most importantly, we uncover an action-angle description of the post-quench dynamics in the thermodynamic limit: this structure naturally explains both the weak equilibration to the GGE and the presence of persistent oscillations, the latter being associated with a resonance at the upper edge of the phonon spectrum. This mechanism provides a physically transparent explanation of weak ergodicity breaking in lattice systems and clarifies why such oscillations disappear in the quantum field theory limit, offering insights expected to remain relevant beyond the specific integrable model considered. To the best of our knowledge, this action-angle description in the thermodynamic limit had not been previously identified. }

\emph{The $O(n)$ Quantum Rotors:} let us consider a $d$-dimensional (cubic) lattice with lattice-spacing $a$ and $N=(L/a)^d$ sites: for each site $\mathbf{j}$ we introduce a $n$-component vector variable $\boldsymbol{\Phi}_{\mathbf{j}}$ and its conjugate  momentum $\boldsymbol{\Pi}_{\mathbf{j}}$, such that $[\Phi^{\alpha}_{\mathbf{j}},\Pi^{\alpha'}_{\mathbf{j'}}] = \text{i} \ \delta_{\alpha,\alpha'} \delta_{\mathbf{j},\mathbf{j}'}$, with $\alpha,\alpha'=1, \dots, n$. The interaction is described by the $O(n)$-symmetric quartic nearest-neighbors Hamiltonian
\begin{figure}
    \centering    \includegraphics[width=\linewidth]{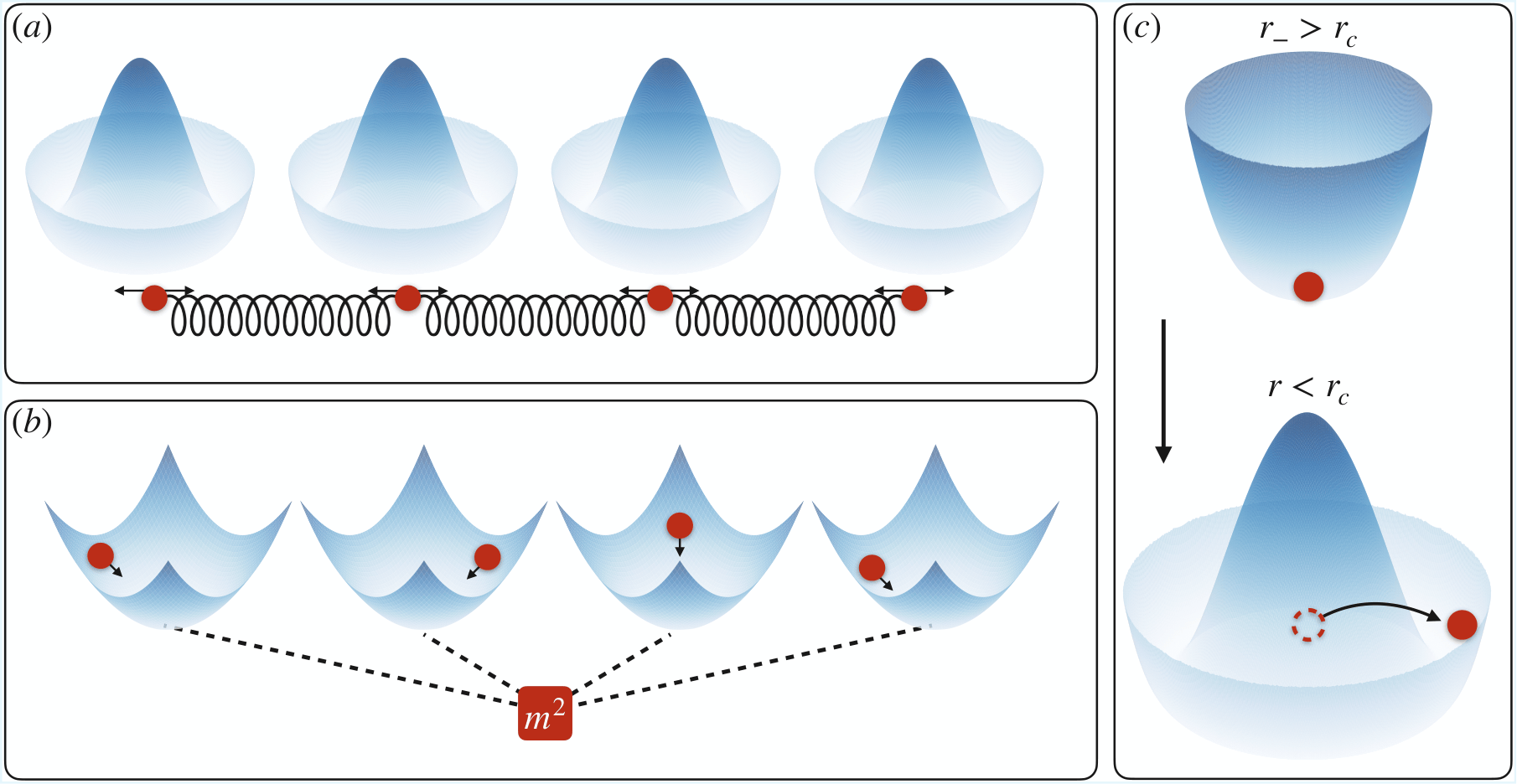}
    \caption{ $(a)$ Sketch of the the quantum $O(n)$ model for $r<0$: the model consist of a chain of locally coupled bosons kept in place by a quartic potential. $(b)$ Sketch of the large-$n$ limit of the model: in this regime the model can be described as gas of phonons scattering through the time-dependent, self consistent mass. $(c)$ Sketch of the quench dynamics: starting from the ground state, the system is driven out of equilibrium by a sudden quench of the bare mass $r^{-} \rightarrow r$ (here $r < 0$).}
   \label{fig: Fig1_schematic}
\end{figure}
\begin{equation}\label{eq:Ham1}
    H = \sum_{\mathbf{j}}   \frac{1}{2} \boldsymbol{\Pi}^2_\mathbf{j} + U(\boldsymbol{\Phi}_\mathbf{j}^2)  + \frac{1}{2a^2} \sum_{\me{\mathbf{j},\mathbf{j'}}} \left( \mathbf{\Phi}_\mathbf{j} - \mathbf{\Phi}_\mathbf{j'} \right)^2,
\end{equation}
where $U(\boldsymbol{\Phi}^2) = r \ \boldsymbol{\Phi}^2/2 + \lambda (\boldsymbol{\Phi}^2)^2/(4n)$ with $\lambda > 0$; $r$ representing the bare mass of the theory, see Fig.\,\ref{fig: Fig1_schematic}a. The Hamiltonian \eqref{eq:Ham1} represents the lattice discretizations of the usual $O(n)$ bosonic field theory, which is recovered as $a \rightarrow 0$, $L \rightarrow \infty$: in particular for $r<0$, depending on the dimensionality , the ground state of model may undergo a transition between a massive ($\me{\boldsymbol{\Phi}} \neq 0$) and a massless ($\me{\boldsymbol{\Phi}} = 0$) phase \cite{sachdev1999quantum}; $\me{\cdot}$ denoting the quantum expectation value. 

The dynamics of the model for a ground state quench is known to become tractable in the limit $n \rightarrow \infty$, due to two facts: $i)$, the ground state of the initial Hamiltonian is homogeneous and factorized over the component indices 
$ii)$ such a factorization is maintained at any finite time $t$~\cite{sotiriadi2010quantum, sciolla2013quantum, chandran2013equilibration}, so that it is possible to replace $\boldsymbol{\Phi}^2_\mathbf{j} \Phi^{\alpha}_\mathbf{j} \rightarrow n \me{\Phi^2_\alpha} \Phi^{\alpha}_\mathbf{j}$ in the Heisenberg equations of motion (e.o.m.) coming from Eq.\,\eqref{eq:Ham1}. 

As a result, in this limit the system can be naturally described in terms of $n$ identical phononic chains, in which each bosonic field $\alpha$ interacts with each others through a collective degree of freedom, which plays the role of an effective, dressed, mass, see Fig.\,\ref{fig: Fig1_schematic}b. In particular, in terms of the Fourier transformed bosons 
$\Phi_\mathbf{k}$, $\Pi_\mathbf{k}$ ($[\Phi_{\mathbf{k}},\Pi_{\mathbf{k'}}] = i \delta_{\mathbf{k},\mathbf{k'}}$), one has
\begin{equation} \label{eq:eomPhiPi}
 \dot{\Phi}_\mathbf{k} =  \Pi_\mathbf{k}^\dagger, \ \ \dot{\Pi}_\mathbf{k} = - (m^2 (t) + \omega_\mathbf{k}^2)  \Phi_\mathbf{k}^\dagger ,
\end{equation}
as the evolution of each component is decoupled, the index $\alpha$ has been dropped. Here, $\omega_\mathbf{k}^2 = 4 a^{-2} \sum_{\nu=1}^d \sin^2 (k_\nu a/2)$ is the lattice dispersion relation, while the effective mass is determined self-consistently by $m^2 (t) = r + \lambda \me{\Phi^2}$, see Fig.\,\ref{fig: Fig1_schematic}. It is worth noting that $m^2(t)$ need not to be positive at any times, while $m^2(t)\geq0$ on any stationary state. 

In the ground-state, the renormalized mass is given by the self consistent relation\,\cite{sotiriadi2010quantum,suppmatt}
\begin{align}
\label{eq:mgs}
m^2_{\rm gs}=r_- +  \frac{\lambda}{2 N} \sum_\mathbf{k} \frac{1}{\sqrt{ m^2_{\rm gs} + \omega_{\mathbf{k}}^2 }} \, .
\end{align}
This allows us to link the quantum critical properties of the model to the infrared behavior of the sum in Eq.~\eqref{eq:mgs}, for $m_{\rm gs} = 0$ in the $N \rightarrow \infty$ limit. If such sum converges, then a (negative) critical value $r^c_{\rm gs}$ of the bare mass $r$ exists such that $m_{\rm gs} \to 0$ as $r \to r^c_{\rm gs}$. This marks the onset of Bose-Einstein condensation in the $\mathbf{k} = 0$ mode. The critical scaling behaviour can be characterized by the scaling of the renormalized mass, $m_{\rm gs} \sim (r - r^c_{\rm gs})^\nu$.
For $d \geq 3$, the system belongs to the mean-field universality class, with a correlation length exponent $\nu = 1/2$. As the dimension is lowered within the range $1 < d < 3$, the exponent acquires a dimensional dependence, $\nu = 1/(d - 1)$, reflecting the increasing influence of fluctuations. In the limit $d \to 1$, the transition disappears as $r^c_{\rm gs} \to 0$~\cite{vojta1996quantum}.

 We will consider the following quench protocol: for $t<0$, we choose $r^{-} > 0$, and assume the system to be in the ground state of the Hamiltonian in Eq.~\eqref{eq:Ham1} with $r=r^{-}$; then at $t=0$ the bare mass is quenched to its (possibly negative) final value $r$, see Fig.~\ref{fig: Fig1_schematic}c. For $t>0$, the evolution of the system is described by Eq.\,\eqref{eq:eomPhiPi}. In turn these can be simplified by parameterizing the second order moments as $|\eta_\mathbf{k} |^2 \equiv \me{\Phi^\dagger_\mathbf{k} \Phi_\mathbf{k}}$, $2 \Re (\eta^{*}_\mathbf{k} p_\mathbf{k}) \equiv \me{\Pi_\mathbf{k} \Phi_\mathbf{k}  + \text{h.c.}}$, $|p_\mathbf{k}|^2 \equiv \me{\Pi^\dagger_\mathbf{k} \Pi_\mathbf{k}}$. Up to normalization, $\eta_\mathbf{k}$, $p_\mathbf{k}$ can also be seen as the components of the expansion of the fields $\Phi_\mathbf{k}$, $\Pi_\mathbf{k}$ in the basis of the ladder operators $a_\mathbf{k}$, $a_\mathbf{-k}$, see e.g. \cite{chandran2013equilibration}. Then, the e.o.m. can be rewritten as
 \begin{align} \label{eq:eompeta}
 \dot{p}_\mathbf{k} = - (m^2 + \omega_\mathbf{k}^2) \eta_\mathbf{k},\qquad
      m^2 = r + \frac{\lambda}{N} \sum_{\mathbf{k}} |\eta_\mathbf{k}|^2 ,
\end{align} 
with  $\dot{\eta}_\mathbf{k} = p_\mathbf{k}$.
Let us notice that, as in our protocol both the e.o.m. and the initial condition only depends on $\omega^2_\mathbf{k}$, so that all the $\eta_\mathbf{k}$ corresponding to degenerate frequencies are in turn degenerate at any $t$. The number of non-degenerate modes will be denoted with $\mathcal{N}$.

\begin{figure*}
    \centering    \includegraphics[width=\linewidth]{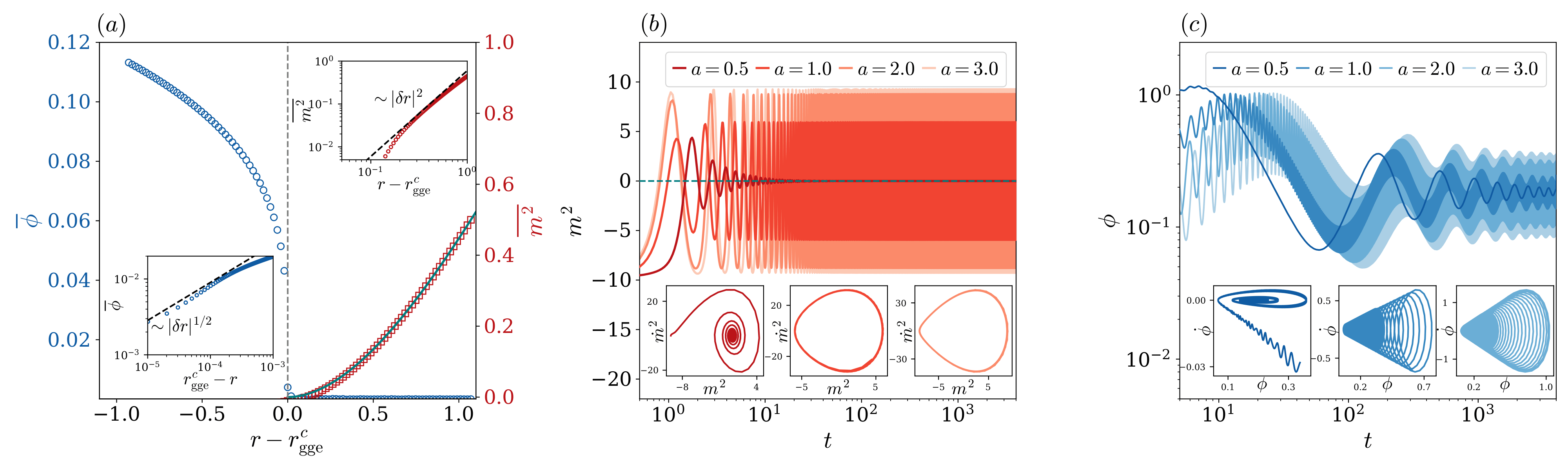}
    \caption{(a) Out-of-equilibrium phase diagram. Blue dots represent the long-time average of the order parameter, $\phi = \eta_0 / \sqrt{N}$ (right $y$-axis), while red squares denote the squared mass, $m^2$ (left $y$-axis), as functions of the final bare mass $r$. For $r < r_\mathrm{gge}^c$, the system exhibits a finite order parameter ($\phi > 0$) and a vanishing mass ($m^2 = 0$), whereas for $r > r_\mathrm{gge}^c$ the order parameter vanishes ($\phi = 0$) and the mass remains finite ($m^2 > 0$). The gray dashed line marks the GGE prediction for the dynamical critical point $r = r_\mathrm{gge}^c$. The numerical results for the averaged squared mass (red squares) are also in excellent agreement with the GGE prediction\,\eqref{eq:mgge} (green solid line). The insets display the scaling behavior of the averaged order parameter (blue dots, left inset) and of the averaged squared mass (red dots, right inset) as functions of $\delta r = |r - r_\mathrm{gge}^c|$ in a log–log scale near the transition. The extracted critical exponents are consistent with the spherical-model universality class in $d = 3$: $\overline{m^2} \sim |\delta r|^{2\nu}$ with $\nu = 1/(d-2) = 1$, and $\overline{\phi} \sim |\delta r|^{\beta}$ with $\beta = 1/2$. Deviations from the power-law scaling of the squared mass at very small $\delta r$ arise from finite-size effects. The system size is $L = 5a\times 10^3$, the lattice spacing is $a = 3$ the initial value of the bare mass is $r^{-} = 1>r_\mathrm{gge}^c$, $\lambda = 1$. (b) Time evolution of the squared mass $m^2(t)$ following a sudden quench from the disordered phase ($r^- = 1$) to the ordered phase ($r = -10$) in $d = 3$, for different lattice spacings $a$ (curves in different shades). The system size is $L = 8a\times 10^3$, with $\lambda = 1$. In the field-theory limit ($a \to 0$), the dynamics converge to the GGE prediction (green dashed line). \textcolor{black}{For any finite lattice spacing, $m^2(t)$ approaches persistent, undamped, oscillations around the GGE prediction. For sufficiently small $a$, however, the asymptotic oscillation amplitude becomes extremely small and is barely visible on the scale of the figure.} (c) Time evolution of the order parameter $\phi(t)$ for the same quench and system parameters as in panel (b). For finite lattice spacing, $\phi(t)$ displays fast, undamped oscillations superimposed to the standard oscillations around the potential minimum in the broken phase. The insets in panels (b) and (c) show the corresponding phase-space trajectories in $(m, \dot{m})$ and $(\phi, \dot{\phi})$ space, respectively.}
   \label{fig: Fig2_PhaseDiagram_m2_phi}
\end{figure*}
\emph{Integrability:} \textcolor{black}{We now show that the semiclassical equations of motion arising in the large-$n$ limit of the quantum $O(n)$ model can be mapped onto a classical integrable Neumann system.} In particular, the e.o.m. in Eq.\,\eqref{eq:eompeta} describe the motion of the classical Hamiltonian 
\begin{equation} \label{eq:classicalH}
\mathcal{H} = \frac{1}{2} \sum_\mathbf{k} \left(|p^2_{\mathbf{k}}| + (r + \omega^2_\mathbf{k}) |\eta^2_{\mathbf{k}}| \right) + \frac{\lambda}{4N} \left( \sum_{\mathbf{k} } |\eta^2_\mathbf{k}| \right)^2   \, .
\end{equation}
As already noticed in Ref.~\cite{chandran2013equilibration}, beside $\mathcal{H}$ itself, the $U(1)$ symmetry of the Hamiltonian 
results in the presence of a set of $N$ trivial integrals of motions, namely $\ell_\mathbf{k} = \Im(\eta^{*}_\mathbf{k} p_\mathbf{k})$: although these can be classically interpreted as angular momenta, at the quantum level their value is fixed by the energy of the quasi-particles in the initial state. Thus, the classical angular momentum  cannot vanish but rather has to obey $\ell_\mathbf{k} \geq 1/2 \quad \forall \mathbf{k}$, due to the quantum zero-point energy \footnote{Notice that for $\mathbf{k} = 0$, the canonical commutation relations fix $\ell_0 = 1/2$.}. As the number of (real) degrees of freedom is $2N$, these conserved quantities are not enough to make the dynamics integrable. 

\textcolor{black}{On the other hand, in the context of classical integrable systems, Hamiltonian~\eqref{eq:classicalH} belongs to the Neumann class of integrable models~\cite{choodnovsky1978completely,wojciechowski1985integrability,neumann1859problemate,neumann1856problemate,uhlenbeck2006equivariant}, whose integrals of motion can be written in the form of dressed single-mode energies}
\begin{equation} \label{eq:iom}
\begin{split}
    &\epsilon_\mathbf{k} = \frac{1}{2} \left(|p^2_{\mathbf{k}}| + (r + \omega^2_\mathbf{k}) |\eta^2_{\mathbf{k}}| \right) + \frac{\lambda}{4N} |\eta^2_{\mathbf{k}}| \sum_{\mathbf{k}' } |\eta^2_\mathbf{k'}| \\
    &+\frac{\lambda}{4N} 
     \sum_{\mathbf{k'}}{\vphantom{\sum}}' \frac{|p^2_{\mathbf{k'}} \eta^2_{\mathbf{k}}| + |p^2_{\mathbf{k}} \eta^2_{\mathbf{k'}}| - 2 \Re{(\eta_\mathbf{k'}^{*} p_\mathbf{k'}) } \Re{(\eta_\mathbf{k}^{*} p_\mathbf{k} } )}{\omega_\mathbf{k'}^2- \omega_\mathbf{k}^2} \, ,
\end{split}
\end{equation}
where the primed sum symbol denotes the summation over all the modes non degenerate with $\omega_{\mathbf{k}}$. Notice that $\mathcal{H} = \sum_{\mathbf{k}} \epsilon_\mathbf{k}$. \textcolor{black}{While the large-$n$ equations of motion were known in the literature, to the best of our knowledge their explicit identification with the classical integrable Neumann system had not been previously recognized. As shown in the following Sections, this correspondence provides direct access to the thermodynamic limit of the integrable dynamics. In particular, it allows us to derive the properties of the generalized Gibbs ensemble and, more importantly, to construct an explicit action-angle description of the thermodynamic dynamics, from which we obtain the exact excitation spectrum. Such an action-angle description of the thermodynamic limit does not appear to have been previously derived in the literature on the classical Neumann model.}

\emph{Generalized Gibbs Ensemble}:
\textcolor{black}{Since the quasiparticle spectrum $\omega_{\mathbf{k}}$ becomes continuous in the thermodynamic limit, it is natural to expect the long-time averages of observables such as $m^2(t)$, $|\eta_{\mathbf{k}}(t)|^2$, and $|p_{\mathbf{k}}(t)|^2$ to converge to the corresponding expectation values $m^2_{\rm gge}$, $|\eta_{\mathbf{k}}|^2_{\rm gge}$, and $|p_{\mathbf{k}}|^2_{\rm gge}$ predicted by the generalized Gibbs ensemble (GGE). Within the GGE, these quantities can be related to the conserved charges $\epsilon_{\mathbf{k}}$ through the virial relation $|p_{\mathbf{k}}|^2_{\rm gge} = (m^2_{\rm gge}+\omega_{\mathbf{k}}^2)|\eta_{\mathbf{k}}|^2_{\rm gge}$, whose validity is proven in the End Matter. Physically, this relation corresponds to the one expected for a gas of non-interacting phonons with effective time-independent mass $m^2_{\rm gge}$. By exploiting this virial relation we can rewrite the $\epsilon_\mathbf{k}$ as}
\begin{equation}
\begin{split}
        \frac{\epsilon_\mathbf{k}}{m^2_{\rm gge} + \omega_\mathbf{k}^2} =  |\eta_\mathbf{k}|^2_{\rm gge} \left( 1 + \frac{\lambda}{2N} \sum_{\mathbf{k'}}{\vphantom{\sum}}' \frac{|\eta_\mathbf{k'}|^2_{\rm gge}}{\omega_\mathbf{k'}^2-\omega_\mathbf{k}^2}\right) 
\end{split}
\end{equation}
up to $O(N^{-1})$ terms which are negligeble if no condensation occurs. We find thus the self consistent condition  
\begin{equation}
    m^2_{\rm gge} = r + \frac{\lambda}{N} \sum_{\mathbf{k}} \frac{\epsilon_\mathbf{k}}{m^2_{\rm gge} + \omega_\mathbf{k}^2}, \label{eq:mgge}
\end{equation}
which agrees with the leading order perturbative result in Refs.~\cite{sotiriadi2010quantum,smacchia2015exploring} as shown in the SM~\cite{suppmatt}.

\emph{Dynamical Critical behavior:} When the sudden quench $r^- \rightarrow r$ from the ground-state is performed, we expect the long-time averaged value $\overline{m^2} = \lim_{T \to \infty} \frac{1}{T} \int_0^T m^2(t)dt$ of the renomalized mass to coincide with its GGE value\,\eqref{eq:mgge}. It can be shown that for a ground state quench, $m_{\rm gge} - m_{\rm gs}$ has the same sign of $r - r_-$, and in particular $m_{\rm gge} = m_{\rm gs}$ iff $r = r_- = r^c_{\rm gs}$. By varying the post-quench bare mass $r$ in the region $r < r_-$ then a critical point $r^{c}_{\rm gge}$ emerges in the GGE equilibrium, such that the renormalized mass $m_{\rm gge}$ in Eq.\,\eqref{eq:mgge} vanishes as $r \to r^{c}_{\rm gge}$. For $r < r^c_{\rm gge}$ the zero mode $\eta_0$ condensates signaling the appearance of a spontaneous magnetization $\phi = \eta_0/\sqrt{N}$.  Around the transition, the system exhibits dynamical scaling behavior, characterized by universal critical exponents $m_{\rm gge} \sim (r-r^c_{\rm gge})^\nu$, $\phi_{\rm gge} \sim (r^c_{\rm gge}-r)^\beta$. From Eq.\,\eqref{eq:mgge}, we find that no symmetry-broken phase exists for dimensions $d \leq 2$, while $\nu = 1/(d - 2)$ for $2 < d < 4$, which reaches the mean-field value $\nu = 1/2$ for $d \geq 4$. The scaling exponent of the order parameter remains dimension independent $\beta=1/2$~\cite{joyce1966spherical,vojta1996quantum}. 

Figure~\ref{fig: Fig2_PhaseDiagram_m2_phi}a presents the dynamical phase diagram of the model in $d=3$ for initial states with $r^{-} > r^c_{\rm gs}$. The solid green line shows the renormalized squared mass $m^2_{\rm gge}$, obtained by solving Eq.~\eqref{eq:mgge} as a function of the post-quench bare mass $r$. The critical point $r^c_{\rm gge}$ is identified by the condition $m^2_{\rm gge} = 0$. The empty squares represent numerical results from time evolution, where the long-time average of the squared mass is computed. These values collapse onto the GGE prediction, supporting the validity of the \emph{kinematical chaos} hypothesis, which asserts that long-time dynamics are governed by statistical ensembles.

The insets in Fig.~\ref{fig: Fig2_PhaseDiagram_m2_phi}a examine the scaling behavior of both the order parameter and the GGE mass. In the vicinity of the critical point, where $|r - r^c| \ll 1$, the scaling exponents match those of the thermal universality class, consistent with the dynamical quantum-to-classical correspondence~\cite{gambassi2011quantum}. For deeper quenches, a crossover to trivial scaling exponents $\nu = 1/2$ and $\beta = 1/4$ emerges, identified in Refs.~\cite{sciolla2013quantum, weidinger2017dynamical}. This crossover can be attributed to the shrinking of the critical scaling window, which is bounded by the condition $m_{\rm gge} \ll 3 m_{\rm gs} a^3 / (8 \pi^2)$ and becomes narrower for small values of $a$~\cite{suppmatt}.

\emph{Weak-ergodicity breaking:} Although the time-averaged squared mass $\overline{m^2}$ matches the GGE prediction, the full post-quench dynamics from the disordered to the ordered phase depends on the lattice spacing $a$. For any finite $a$, $m(t)$ exhibits persistent, undamped oscillations around the GGE value, signaling weak ergodicity breaking. As shown in Fig.~\ref{fig: Fig2_PhaseDiagram_m2_phi}b, these oscillations become asymptotically periodic, with their amplitude vanishing as $a = \Lambda^{-1} \to 0$. This behavior stems from a resonance at the edge of the quasi-particle spectrum and persists even in the thermodynamic limit ($N \to \infty$), challenging the expectations of the \emph{kinematical chaos} hypothesis.

\begin{figure}
    \centering    \includegraphics[width=\linewidth]{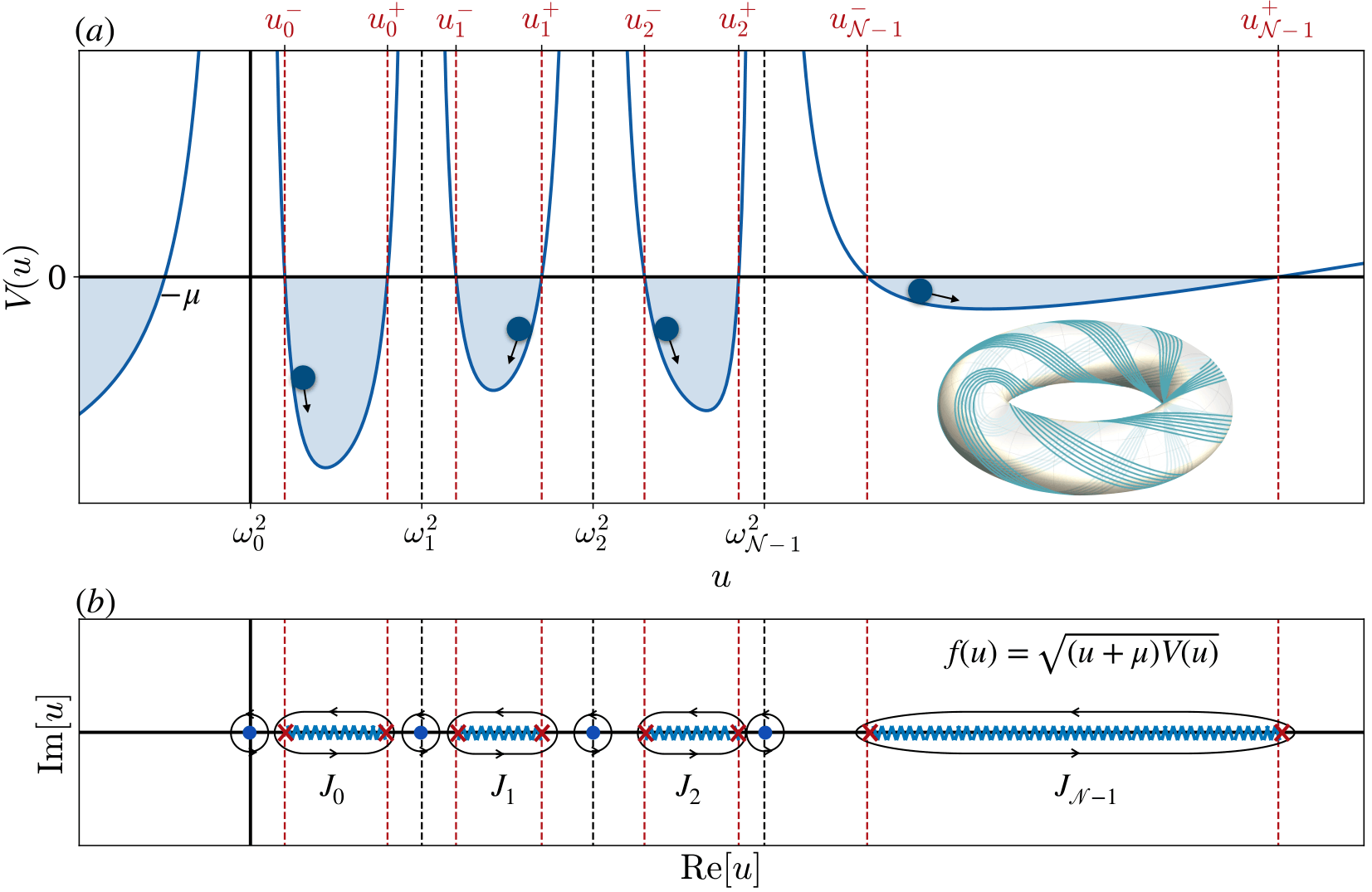}
    \caption{$(a)$ Jacobi quasiparticle picture: each microscopic degree of freedom oscillates between the inversion points $u_l^{\pm}$ of the potential $V(u)$ for $u >0$. The negative zero $u =-\mu$ of $V(u)$ physically represent the effective GGE mass $m_{\rm gge}^2$. As the $u_l^{\pm}$ are bounded by the bare frequencies $\omega_l^2$, as the spectrum becomes continuous for $N \rightarrow \infty$, the first $\mathcal{N}-1$ Jacobi quasiparticles acquire a well-defined energy, while the oscillation $\mathcal{N}$-th quasiparticle beyond the edge of the spectral band has a finite support, leading to the emergence of a Higgs mode with persistent oscillations. 
    $(b)$ The set of action variables $\lbrace J_l, \ell_l \rbrace$ which constraints the dynamics on invariant tori (see inset) can be computed from the complex plane singularities of $f(u) = \sqrt{(u+\mu) V(u)}$, as integrals around its branch cuts and residues at the poles $\lbrace \omega^2_l \rbrace$ respectively. In the QFT limit ($a \rightarrow 0$), this allows the direct computation of the spectrum of the model and a proof of the thermalization to the GGE predictions.
    }
   \label{fig: Fig3_schematic}
\end{figure}

This resonance phenomenon is better understood in terms of Jacobi quasiparticles. The Jacobi coordinates $u_{l}$ and momenta $P_{l}$ are a set of separable coordinates~\cite{wojciechowski1985integrability} for the Hamiltonian~\eqref{eq:classicalH}, their motion is constrained to the zero energy subspace of the effective Hamiltonian $\mathcal{H}_{J}=\lambda^2 P_l^2/N^2 + V(u_l) = 0$, where $V(u)$ is the potential
\begin{equation} \label{eq:V(u)}
    V(u) = u + r - \frac{\lambda}{N} \sum_{\mathbf{k}} \frac{\epsilon_\mathbf{k}}{u-\omega^2_\mathbf{k}} + \frac{\lambda^2}{4N^2}\sum_\mathbf{k} \frac{\ell_\mathbf{k}^2}{(u-\omega^2_\mathbf{k})^2}  .
\end{equation}
Since the motion is confined to the region where $V(u) < 0$ and $u > 0$, the positive zeros of $V(u)$ define the turning points of each coordinate $u_l$, denoted as $u_l^{\pm}$. The quantum condition $\ell_\mathbf{k} \geq 1/2$ ensures these coordinates remain bounded. For $l \neq \mathcal{N} - 1$, the turning points satisfy $\omega^2_l < u_l^{-} < u_l^{+} < \omega^2_{l+1}$, while the final pair $u_{\mathcal{N}-1}^{\pm}$ lies beyond the band edge and is only bounded from below by the ultraviolet cutoff $\omega^2_{\mathcal{N}-1} \equiv \omega_\Lambda^2$, as shown in Fig.~\ref{fig: Fig3_schematic}a. The potential also has a negative zero at $u = -\mu < 0$. In the thermodynamic limit ($N \to \infty$), the spectrum becomes continuous $|\omega_{l}-\omega_{l+1}|\approx 1/N$ and the quasiparticles $u_l$ for $l = 0, \dots, \mathcal{N}-2$ localize around well-defined energies. However, the behavior of the final quasiparticle $u_{\mathcal{N}-1}$ depends on whether the limit is taken at finite or vanishing lattice spacing.

\emph{The quantum field theory limit:} In the limit where the lattice spacing vanishes ($a \to 0$), the system approaches the quantum field theory (QFT) regime, and the excitation spectrum $\lbrace\Omega_{\mathbf{k}}\rbrace$ can be computed explicitly by analyzing the singularity structure of the auxiliary function $f(u) = \sqrt{(u + \mu) V(u)}$ in the complex plane. In this continuum limit, contour integrals around the branch cuts of $f(u)$ can be directly related to the classical action variables via $J_l = \oint \frac{du_l}{2\pi} , P_l(u_l)$, while the residues at the poles located at $\omega_l^2$ correspond to the $\ell_l$, as illustrated in Fig.~\ref{fig: Fig3_schematic}b. The explicit expression for the excitation frequencies is derived in the SM~\cite{suppmatt}
\begin{equation} \label{eq:Omegal}
    \Omega_{\mathbf{k}} = \sqrt{\mu + \omega_\mathbf{k}^2} \, .
\end{equation} 
Outside the bandwidth region, where $u \notin [0, \Lambda]$, the last term in Eq.~\eqref{eq:V(u)} becomes negligible, scaling as $O(N^{-1})$. As a result, in the thermodynamic limit, the parameter $\mu$ coincides with the GGE mass $m_{\rm gge}^2$ defined in Eq.~\eqref{eq:mgge}. This allows Eq.~\eqref{eq:Omegal} to be interpreted as a gas of free bosons with mass $m_{\rm gge}$. Therefore, the QFT regime adheres strictly to the \emph{kinematical chaos hypothesis}~\cite{lasinio1996chaotic}, which posits strong equilibration to the GGE.

Also for finite $a$ the phonon dispersion $\omega_l^2$ becomes continuous in the thermodynamic limit. Then, the width of each oscillation, $u_l^{+} - u_l^{-}$ for $l = 0, \dots, \mathcal{N}-2$, scales as $O(N^{-1})$. However, the final mode $u_{\mathcal{N}-1}$ lies beyond the band edge $\Lambda$ and retains a finite support, resulting in an isolated frequency that persists in the continuum limit $N\to\infty$ (see SM\,\cite{suppmatt} for the details). This isolated mode is responsible for long-lived oscillations in the system. In fact, the contribution from $u_{\mathcal{N}-1}$ leads to undamped periodic oscillations of observables with a characteristic frequency $\omega_{\mathcal{N}-1}$, surviving the dephasing process over long times. As explicitly demonstrated in Ref.~\cite{collado2023engineering}, similar syncronization phenomena arise due to singularities in the quasi-particle spectrum and are, therefore, absent in the case of smooth, e.g Gaussian or exponential, regularization \cite{maraga2015aging,chiocchetta2015shorttime}.

\emph{Conclusions:} In this Letter, we identify a general dynamical mechanism underlying weak ergodicity breaking after quantum quenches: a resonance at the upper edge of the quasi-particle band. Using the large-$n$ limit of the $O(n)$ symmetric model as a paradigmatic and exactly solvable setting, we show that persistent oscillations are a high-energy, lattice-induced phenomenon, which vanish as the lattice spacing $a$—or equivalently, the ultraviolet cutoff $\Lambda$—is removed. The relevance of this mechanism to ergodicity breaking observed in numerical simulations~\cite{banuls2011strong,rakovszky2016hamiltonian,robinson2019signatures} can be tested by analyzing how the amplitude and frequency of oscillations depend on the lattice spacing.

Within the same framework, we provide an explicit demonstration of GGE thermalization in this model, resolving the puzzle posed in Ref.~\cite{chandran2013equilibration}. In the quantum field theory limit ($N,\Lambda\to\infty$), the relevant observables equilibrate to the GGE in a strong sense, while for finite lattice spacing they display persistent oscillations around the GGE averages [Fig.~\ref{fig: Fig2_PhaseDiagram_m2_phi}(b,c)]. We further identify the dynamical critical scaling regime and show that it belongs to the thermal universality class, in agreement with the quantum-to-classical correspondence~\cite{gambassi2011quantum}.

An important open question concerns the fate of persistent oscillations beyond the integrable limit. While perturbative approaches suggest undamped oscillations even at long times~\cite{delfino2020persistent,delfino2022persistent,sorba2025quantum}, other studies predict a finite damping that increases with quench strength~\cite{robertson2024decay}. In our case, oscillations are protected by integrability, but incorporating $1/N$ corrections is expected to clarify their stability and may enable the exploration of more general dynamical critical phenomena, such as symmetry re-breaking recently observed in $\phi^4$ theories~\cite{balducci2025symmetry}.

\emph{Data Availability}: scripts and data used in this work are available at \footnote{\url{https://github.com/andrea-solfanelli/Universality-and-weak-ergodicity-breaking-in-quantum-quenches}}.

\emph{Acknowledgements}: GG acknowledges the support of the MSCA Grant 101152898 (DREAMS). This research was funded by the Swiss National Science Foundation (SNSF) grant numbers 200021--207537 and 200021--236722, by the Deutsche Forschungsgemeinschaft (DFG, German Research Foundation) under Germany's Excellence Strategy EXC2181/1-390900948 (the Heidelberg STRUCTURES Excellence Cluster) and by the European Union under GA No. 101077500–QLRNet. Partial support by grant NSF PHY-230935 to the Kavli Institute for
Theoretical Physics (KITP) is also acknowledged.

%

\section{End Matter}
\subsection{GGE ensemble}
In this section we will analyze the properties of the GGE equilibrium ensemble, namely 
\begin{equation} \label{eq:GGE}
    P(p,\eta) = \frac{1}{Z} e^{- \sum_{\mathbf{k}} \beta_{k} \epsilon_{\mathbf{k}} (p,\eta) + \varpi_\mathbf{k} \ell_\mathbf{k} (p,\eta)} ,  
\end{equation}
where the Lagrange multipliers $\beta_\mathbf{k}$,$\varpi_{\mathbf{k}}$ are fixed by the values of the integral of motion; and in particular we will show that the virial condition $|p^2_\mathbf{k}|_{\rm gge} = (m^2_{\rm gge} + \omega_\mathbf{k}^2) |\eta^2_\mathbf{k}|_{\rm gge}$ actually holds, where $A(p,\eta)_{\rm gge}$ represent the average of the function $A(p,\eta)$ over the probability distribution \eqref{eq:GGE}. 

We start by rescaling the momenta by a factor $p_{\mathbf{k}} = \sqrt{\mu^2 + \omega_\mathbf{k}^2} \ \tilde{p}_\mathbf{k}$ for some $\mu > 0$. We find that 
\begin{equation}
    \begin{split}
        \ell_\mathbf{k} (\tilde{p},\eta) &= \sqrt{\mu + \omega_\mathbf{k}^2} \ \Im(\eta_\mathbf{k}^{*} \tilde{p}_\mathbf{k}) \\ 
        \epsilon_\mathbf{k} (\tilde{p},\eta) &= \frac{1}{2} (\mu + \omega_\mathbf{k}^2)  \left( |\tilde{p}^2_\mathbf{k}| + |\eta^2_\mathbf{k}| \vphantom{\frac{1}{2}} + \frac{\lambda}{2N} \times \   \right. \\ &\left.   \sum_{\mathbf{k'}}{\vphantom{\sum}}' \frac{|\tilde{p}^2_{\mathbf{k'}} \eta^2_{\mathbf{k}}| + |\tilde{p}^2_{\mathbf{k}} \eta^2_{\mathbf{k'}}| - 2 \alpha_{\mathbf{k},\mathbf{k}'} \Re{(\eta_\mathbf{k'}^{*} \tilde{p}_\mathbf{k'}) } \Re{(\eta_\mathbf{k}^{*} \tilde{p}_\mathbf{k} } )}{\omega_\mathbf{k'}^2- \omega_\mathbf{k}^2} \right) \\
    &+ \frac{1}{2}\left( \mu - r - \frac{\lambda}{2N} \sum_{\mathbf{k'}} (|\eta_\mathbf{k'}^2| + |\tilde{p}_\mathbf{k'}^2|) \right) |\eta_\mathbf{k}^2| \, , 
    \end{split}
\end{equation} 
with $\alpha_{\mathbf{k},\mathbf{k}'} = \sqrt{(\mu + \omega_\mathbf{k'}^2)/(\mu + \omega_\mathbf{k}^2)}$. For $N \rightarrow \infty$, the fluctuations of the quantity within brackets in the last line are going to be suppressed, so that we can replace it with its ensemble average. By choosing thus
\begin{equation}
    \mu = r + \frac{\lambda}{2N} \sum_{\mathbf{k'}} (|\eta_\mathbf{k'}^2|_{\rm gge} + |\tilde{p}_\mathbf{k'}^2|_{\rm gge}) \, , 
\end{equation}
we have that the probability measure becomes symmetric under the exchange $\eta_\mathbf{k} \leftrightarrow \tilde{p}_\mathbf{k}$. As a consequence $|\tilde{p}_\mathbf{k'}^2|_{\rm gge} = |\eta_\mathbf{k'}^2|_{\rm gge}$ and 
\begin{equation}
    \mu = r + \frac{\lambda}{N} \sum_{\mathbf{k'}} |\eta_\mathbf{k'}^2|_{\rm gge} = m^2_{\rm gge}
\end{equation}
so that, finally, coming back to the original variables, one has the virial condition $|p^2_\mathbf{k}|_{\rm gge} = (m^2_{\rm gge} + \omega_\mathbf{k}^2) |\eta^2_\mathbf{k}|_{\rm gge}$.

\subsection{Jacobi coordinates}
By setting $\eta_{\mathbf{k}} = \xi_\mathbf{k} e^{i \theta_\mathbf{k}}$, we have that $\theta_\mathbf{k}$, $\ell_\mathbf{k}$ are trivially a pair of separable canonical coordinates, as $\theta_\mathbf{k}$ is cyclic in $\mathcal{H}$. We will now prove that the Jacobi coordinates, defined by the relation 
\begin{equation} 
    \varphi(u) = \ 1 - \frac{\lambda}{2N} \sum_{l} \frac{\xi_\mathbf{k}^2}{u-\omega_\mathbf{k}^2}  = \prod_{l} \frac{u-u_l}{u-\omega_l} \, , 
\end{equation}
form a set of separable coordinates for the Hamiltonian $\mathcal{H}$ \eqref{eq:classicalH} (notice that in the last equality we chose to label the non-degenerate frequencies such that $\omega_{l+1} > \omega_l$). By deriving it twice with respect to time and by using the Eqs.\,\eqref{eq:eomPhiPi} together with the identity 
\begin{equation}
    \frac{1}{(u-u_l)(u-u_{l'})} = \frac{1}{u_l-u_{l'}} \left( \frac{1}{u-u_l} - \frac{1}{u-u_{l'}} \right)
\end{equation}
(valid for $l \neq l^\prime$) we find 
\begin{equation}
    V(u) = (u+m^2) \varphi^2 + \frac{1}{2} \varphi \ \partial^2_t \varphi(u) - \frac{1}{4} (\partial_t \varphi)^2 \ . 
\end{equation}
From this we find that, in correspondence of $u = u_l$, one has
\begin{equation}
     V(u_l) + \frac{1}{4} (\partial_t \varphi)^2|_{u=u_l} = 0 \, .
\end{equation}
To prove the separability, we have now to prove that the conjugate momentum to $u_l$ is $P_l = N/(2 \lambda) \ \partial_t \varphi|_{u=u_l}$. To see that, it is convenient to write explicitly the $\xi_l$ in terms of the $u_l$, by expressing the residues of the poles of $\varphi(u)$ in terms of the $u_l$, namely 
\begin{equation}
    \frac{\lambda}{4N} \xi_l^2 = (\omega_l^2 - u_l) \prod_{\omega_l \neq \omega_{l'}} \frac{\omega_{l}^2 - u_{l'}}{\omega_{l}^2-\omega_{l'}^2} \, , 
\end{equation}
from which
\begin{equation}
    \frac{1}{\xi_l} \frac{\partial \xi_{l}}{\partial u_{l'}} = \frac{1}{2} \frac{1}{\omega_l^2-u_{l'}} \, . 
\end{equation}
We find then the momentum $P_l$ corresponding to $u_l$ is given by 
\begin{equation}
    P_l = \sum_{l'} \frac{\partial \xi_{l'}}{\partial u_{l}} p_{l'} = \frac{1}{2} \sum_{l'} \frac{\xi_{l'} p_{l'}}{\omega_{l'}^2 - u_l} = - \frac{N}{2 \lambda} \partial_t \varphi|_{u = u_l} 
\end{equation}
where we used the equation of motion $\dot{\xi}_l = p_l$ for the original variables. 

Let us notice that, more explicitly, one has
\begin{equation} \label{eq:p(qdot)}
    P_l =  \frac{N}{2\lambda}  \frac{ \dot{u}_l}{u_l-\omega_{l}^2}  \prod_{k^\prime \neq k} \frac{u_l - u_{l^\prime}}{u_l - \omega_{l^\prime}^2} \, . 
\end{equation} 

\begin{widetext}
\newpage
\begin{center}
\textbf{\Large Supplemental Material: Integrable Dynamics and Thermalization in the Quantum $O(n)$ Model at Large $n$}
\end{center}

\section{Integrals of Motion}
\noindent
We will now show that the quantities $\epsilon_\mathbf{k}$ defined in Eq.\,(5), namely 
\begin{equation} 
\begin{split}
    \epsilon_\mathbf{k} = \frac{1}{2} \left(|p^2_{\mathbf{k}}| + (r + \omega^2_\mathbf{k}) |\eta^2_{\mathbf{k}}| \right) + \frac{\lambda}{4N} |\eta^2_{\mathbf{k}}| \sum_{\mathbf{k}' } |\eta^2_\mathbf{k'}| +\frac{\lambda}{4N} 
     \sum_{\mathbf{k'}}{\vphantom{\sum}}' \frac{|p^2_{\mathbf{k'}} \eta^2_{\mathbf{k}}| + |p^2_{\mathbf{k}} \eta^2_{\mathbf{k'}}| - 2 \Re{(\eta_\mathbf{k'}^{*} p_\mathbf{k'}) } \Re{(\eta_\mathbf{k}^{*} p_\mathbf{k} } )}{\omega_\mathbf{k'}^2- \omega_\mathbf{k}^2} ,
\end{split}
\end{equation}
are actually integrals of motions for the Hamiltonian evolution
\begin{equation}
    \dot{\eta}_\mathbf{k} = p_\mathbf{k}, \hspace{1cm} \dot{p}_\mathbf{k} = - (m^2 + \omega_\mathbf{k}) \eta_\mathbf{k}
\end{equation} 
with $m^2 = 1 + \lambda/N \sum_{\mathbf{k}} |\eta_\mathbf{k}^2|$. First, it is convenient to rewrite the $\epsilon_\mathbf{k}$ as
\begin{equation} \label{eq:epskVTQ}
    \epsilon_\mathbf{k} = \frac{1}{2} \begin{pmatrix} p_\mathbf{k}^*, \eta_\mathbf{k}^* \end{pmatrix} \begin{pmatrix}
        U_\mathbf{k} & - Q_\mathbf{k} \\ -Q_\mathbf{k} & T_\mathbf{k}
    \end{pmatrix} \begin{pmatrix} p_\mathbf{k} \\ \eta_\mathbf{k} \end{pmatrix} 
\end{equation}
where
\begin{equation} 
\begin{split}
     U_\mathbf{k} &= 1 + \frac{\lambda}{2 N} \sum_\mathbf{k'}{\vphantom{\sum}}' \frac{|\eta_\mathbf{k'}|^2}{\omega^2_\mathbf{k'} - \omega^2_\mathbf{k} }, \\ T_\mathbf{k} &= r + \omega^2_\mathbf{k} + \frac{\lambda}{2 N} \sum_\mathbf{k'}{\vphantom{\sum}}' \frac{|p_\mathbf{k'}|^2}{\omega^2_\mathbf{k'} - \omega^2_\mathbf{k} } + \frac{\lambda}{2 N} \sum_{\mathbf{k'}} |\eta_\mathbf{k'}|^2, \\ Q_\mathbf{k} &=  \frac{\lambda}{2 N} \sum_\mathbf{k'}{\vphantom{\sum}}' \frac{\Re \left( \eta^{*}_\mathbf{k'} p_\mathbf{k'} \right)}{\omega^2_\mathbf{k'} - \omega^2_\mathbf{k} } \, .
\end{split}
\end{equation}
The evolution of $U_\mathbf{k}$,$Q_\mathbf{k}$,$T_\mathbf{k}$ is described by a similar set of closed ODEs, namely 
\begin{equation} \label{eq:RTQdot}
    \dot{U}_\mathbf{k} = 2 Q_\mathbf{k},  \ \ \dot{T}_\mathbf{k} = - 2 (m^2 + \omega^2_k)  Q_\mathbf{k} + \frac{\lambda \nu_\mathbf{k}}{2N} \Re(\eta_\mathbf{k}^* p_\mathbf{k}) , \ \ \dot{Q}_\mathbf{k} = T_\mathbf{k} - (m^2 + \omega^2_\mathbf{k}) U_\mathbf{k} +\frac{\lambda \nu_\mathbf{k}}{2N} |\eta_\mathbf{k}^2| \ , 
\end{equation}
$\nu_\mathbf{k}$ being the degeneracy of the mode $\omega_\mathbf{k}$, so that 
\begin{equation}
\begin{split}
    \dot{\epsilon}_\mathbf{k} = &\frac{1}{2} {\begin{pmatrix} p_\mathbf{k}^*, \eta_\mathbf{k}^* \end{pmatrix}} {\begin{pmatrix}
        \dot{U}_\mathbf{k} - 2 Q_\mathbf{k} & -\dot{Q}_\mathbf{k} - (m^2+\omega_\mathbf{k}^2) U_\mathbf{k} + T_\mathbf{k} \\ -\dot{Q}_\mathbf{k} - (m^2+\omega_\mathbf{k}^2) U_\mathbf{k} + T_\mathbf{k} & \dot{T}_\mathbf{k} + 2 (m^2 + \omega_\mathbf{k}^2)
    \end{pmatrix}} {\begin{pmatrix} p_\mathbf{k} \\ \eta_\mathbf{k} \end{pmatrix}}  \\ 
    =& \frac{\lambda \nu_\mathbf{k}}{4N} {\begin{pmatrix} p_\mathbf{k}^*, \eta_\mathbf{k}^* \end{pmatrix}} {\begin{pmatrix}
        0 & - |\eta_\mathbf{k}^2| \\ - |\eta_\mathbf{k}^2| & 2\Re(\eta^*_\mathbf{k} p_\mathbf{k})
    \end{pmatrix}} {\begin{pmatrix} p_\mathbf{k} \\ \eta_\mathbf{k} \end{pmatrix}} =  0 \, . 
\end{split}
\end{equation}

\section{GGE EQUILIBRIUM}
By exploiting the virial relation $|p_\mathbf{k}^2|_{\rm gge} = (m^2_{\rm gge} + \omega_\mathbf{k}^2) |\eta_\mathbf{k}^2|_{\rm gge}$, proven in tha main text, the definition of the integral of motions gives
\begin{equation} \label{eq:eketagge}
\begin{split}
     \epsilon_\mathbf{k} &= (m^2_{\rm gge} + \omega_\mathbf{k}^2) |\eta_\mathbf{k}|^2_{\rm gge} \left( 1 + \frac{\lambda}{2N} \sum_{\mathbf{k'}}{\vphantom{\sum}}' \frac{|\eta_\mathbf{k'}|^2_{\rm gge}}{\omega_\mathbf{k'}^2-\omega_\mathbf{k}^2}\right) - \frac{\lambda}{4N}|\eta^4_\mathbf{k}|_{\rm gge}. 
\end{split}
\end{equation}
In the following we will assume $m_{\rm gge} >0 $ so that the $|\eta_\mathbf{k}^2|_{\rm gge} = O(1)$ and the last term can be neglected for $N \rightarrow \infty$. In this case, by defining both sides by $(m^2_{\rm gge} + \omega_\mathbf{k}^2) $ and summing over the momenta we get Eq.\,(8) of the main text:
\begin{equation}
    m^2_{\rm gge} = r + \frac{\lambda}{N} \sum_{\mathbf{k}} \frac{\epsilon_\mathbf{k}}{m^2_{\rm gge} + \omega_\mathbf{k}^2} \, .
\end{equation}

\subsection{Mode distribution}
It is also possible to derive an expression for the $|\eta_\mathbf{k}^2|_{\rm gge}$ by inverting Eq.\,\eqref{eq:eketagge}. Indeed for any $u$ far from the spectral band $\lbrace \omega_\mathbf{k}^2 \rbrace$ (i.e. $u \notin [0,\Lambda]$ for large $N$) one finds
\begin{equation}
    \frac{\lambda}{N} \sum_\mathbf{k} \frac{\epsilon_\mathbf{k}}{m^2_{\rm gge}+\omega_\mathbf{k}^2} \frac{1}{u- \omega_\mathbf{k}^2} = 1 - \left( 1 - \frac{\lambda}{2N} \sum_{\mathbf{k}} \frac{|\eta_\mathbf{k}|^2_{\rm gge}}{u-\omega_\mathbf{k}^2} \right)^2 \equiv 1 - \varphi_{\rm gge}^2 (u) \, .
\end{equation}
In the $N \rightarrow \infty$ the spectral band becomes a continuum $[0, \Lambda]$ so that the above expression can be written in integral form as 
\begin{equation}
     \frac{\lambda}{2} \int_0^\Lambda dE \frac{\rho(E) |\eta^2(E)|_{\rm gge}}{u-E} = 1  - \sqrt{1 - \lambda \int_0^\Lambda \frac{dE}{E+m_{\rm gge}^2} \frac{\rho(E) \epsilon(E)}{u-E} } \, ,
\end{equation}
$\rho(E)$ being the density of states of the phononic dispersion $\omega_\mathbf{k}^2$. Finally, by taking the analytic continuation of the equality above just above the branch cut $[0, \Lambda]$ and using the Plemelj formula, one finds 
\begin{equation}
    \frac{\lambda \pi}{2} \rho(E) \ |\eta^2(E)|_{\rm gge} =  \Im \sqrt{1 - \lambda \int_0^\Lambda \frac{dE'}{ E'+m_{\rm gge}^2} \frac{\rho(E') \epsilon(E')}{ E-E' + i 0}} \, .
\end{equation}

\section*{Ground state quench}
\subsection*{Ground state properties}
For $t=0$ the system is in its ground state with $r(t=0)=r_-$. In this state, every mode is in the ground state of a two-dimensional harmonic oscillator of frequency $\Omega_{\mathbf{k},-}^2 =  m^2_{\rm gs}+ \omega_\mathbf{k}^2$, $m^2_{\rm gs}$ being the value of effective mass over the ground state. Since $m^2_{\rm gs}$ is now constant the EoMs can be solved in terms of a set of two ladder operators $a_\mathbf{k}$, $a_{-\textbf{k}}$
\begin{equation}
\Phi_\mathbf{k}  (t) = \frac{1}{\sqrt{2 \Omega_{k,-}}} \left(e^{i \Omega_{\mathbf{k},-} t} a_{\mathbf{k},+} + e^{-i \Omega_{\mathbf{k},-} t} a^\dagger_{\mathbf{k},-} \right) \hspace{1cm} \Pi_\mathbf{k}  (t) = -i \sqrt{\frac{\Omega_{\mathbf{k},-}}{2}} \left(e^{i \Omega_{\mathbf{k},-} t}a^\dagger_{\mathbf{k},+} - e^{-i \Omega_{\mathbf{k},-} t} a_{\mathbf{k},-} \right) 
\end{equation}
so that
\begin{equation}
\begin{split}
|\eta_\mathbf{k} (t)|^2 &= \me{\Phi_\mathbf{k} (t) \Phi_{\mathbf{k}}(t)^{\dagger}} = \frac{1}{2 \Omega_{\mathbf{k},-}} = \frac{1}{2\sqrt{m_{\rm gs}^2 + \omega_\mathbf{k}^2}} \\ 
|p_\mathbf{k}^2 (t)| &=  \me{\Pi_\mathbf{k} (t) \Pi_{\mathbf{k}}(t)^{\dagger}} = \frac{1}{2} \Omega_{\mathbf{k},-} = \frac{1}{2} \sqrt{m_{\rm gs}^2 + \omega_\mathbf{k}^2} \\
\eta^{*}_\mathbf{k} (t) p_\mathbf{k} (t) &= \me{\Phi_\mathbf{k} (t) \Pi_{\mathbf{k}}(t)} = \frac{\text{i}}{2} \ ,\label{SM: intial conditions}      
\end{split}
\end{equation}
and $m^2_{gs}$ is determined by the the self consistency condition
\begin{equation} \label{SM:mugs}
m^2_{\rm gs} = r_- + \frac{\lambda}{2N} \sum_k \Omega_{\mathbf{k},-}^{-1} = r_- +  \frac{\lambda}{2 N} \sum_\mathbf{k} \frac{1}{\sqrt{ m^2_{\rm gs} + \omega_{\mathbf{k}}^2 }} \, .
\end{equation}
so that the ground-state critical point takes place for $r_- = r^c_{\rm gs}$, with 
\begin{equation}
    r^c_{\rm gs} = - \frac{\lambda}{2N} \sum_\mathbf{k} \omega_\mathbf{k}^{-1} \,.\label{SM:r_c_gs}
\end{equation}
Let us now consider the quench $r_- \rightarrow r$: in this case we have
\begin{equation}
\begin{split}
    \ell_\mathbf{k} = \frac{1}{2} \hspace{1cm}
     \epsilon_\mathbf{k} 
     = \frac{1}{2} \sqrt{m^2_{\rm gs} + \omega_\mathbf{k}^2} + \frac{r - r_-}{4 \sqrt{m^2_{\rm gs} + \omega_\mathbf{k}^2} }  + \frac{\lambda}{8N}  \sum_{\mathbf{k'}}{\vphantom{\sum}}^\prime \frac{1}{\omega_{\mathbf{k}^\prime}^2 - \omega_\mathbf{k}^2} \sqrt{\frac{m^2_{\rm gs} + \omega_\mathbf{k}^2 }{m^2_{\rm gs} + \omega_\mathbf{k'}^2} } \, . 
\end{split}
\end{equation}

\subsection{Critical properties for a Ground state quench}
The GGE prediction for $m_{gge}$, given by Eq. (8) of the main text, 
\begin{equation}
    m_{\rm gge}^2 = r + \frac{\lambda}{N} \sum_\mathbf{k} \frac{\epsilon_\mathbf{k}}{m^2_{\rm gge} + \omega_\mathbf{k}^2} ,
\end{equation}
reads for a ground-state quench as 
\begin{equation} \label{SM:muggemugs}
\begin{split} 
    m_{\rm gge}^2 = r &+ \frac{\lambda}{2N} \sum_\mathbf{k} \frac{\sqrt{m^2_{\rm gs} + \omega_\mathbf{k}^2 }}{m_{\rm gge}^2+ \omega_\mathbf{k}^2} +  (r-r_-) \ \frac{\lambda}{4N} \sum_\mathbf{k} \frac{1}{(m_{\rm gge}^2 + \omega_\mathbf{k}^2)\sqrt{m_{\rm gs}^2 + \omega_\mathbf{k}^2}} \\ &+ (m_{\rm gs}^2-m^2_{\rm gge}) \left( \frac{\lambda}{4N} \sum_\mathbf{k} \frac{1}{(m_{\rm gge}^2 + \omega_\mathbf{k}^2)\sqrt{m_{\rm gs}^2 + \omega_\mathbf{k}^2}} \right)^2 \, .
    \end{split}
\end{equation}
Together with Eq.\,\eqref{SM:mugs}, which links $r_-$ and $m_{\rm gs}$, the above Eq.\,\eqref{SM:muggemugs} fixes the value of $m_{\rm gge}$ in terms of $r_-$ (and thus $m_{\rm gs}$). In particular, with the help of Eq.\,\eqref{SM:mugs} we find
\begin{equation} \label{eq:rmgge}
    r - r_- = (m_{\rm gge}^2-m^2_{\rm gs}) \left( 1 + \frac{\lambda}{4N} \sum_\mathbf{k} \frac{1}{(m_{\rm gge}^2 + \omega_\mathbf{k}^2)\sqrt{m_{\rm gs}^2 + \omega_\mathbf{k}^2}} \right) \, 
\end{equation}
from which we find that $m_{\rm gge} = m_{\rm gs}$ iff $r = r^-$. Moreover, we find that for a quench from the critical ground-state ($r_- =r^c_{\rm gs}$, $m_{\rm gs} = 0$) we have that the GGE critical point coincide with the thermal one.  

From Eq.\,\eqref{eq:rmgge}, setting $m_{\rm gge}=0$, we find
\begin{equation}
    r^c_{\rm gge} = r_- - m^2_{\rm gs}  \left( 1 + \frac{\lambda}{4N} \sum_\mathbf{k} \frac{1}{ \omega_\mathbf{k}^2\sqrt{m_{\rm gs}^2 + \omega_\mathbf{k}^2}} \right) = -\frac{\lambda}{2N} \sum_\mathbf{k} \frac{1}{\sqrt{m_{\rm gs}^2 + \omega_\mathbf{k}^2}} \left( 1 + \frac{m^2_{\rm gs}}{2 \omega_\mathbf{k}^2}  \right)
\end{equation}
As the expression above is surplisingly linear in $\lambda$, the exact result coincide with the first order perturbative expansion obtained in Refs. \cite{sotiriadi2010quantum,smacchia2015exploring}.

By expanding close to the critical point we have 
\begin{equation}
    r - r^c_{\rm gge} \sim m_{\rm gge}^{d-2} + O(m_{\rm gge}^2), 
\end{equation}
so that $\nu= (d-2)^{-1}$ for $2 <d < 4$. In particular for $d=3$ one can compute the series coefficients within the Debye approximation finding 
\begin{equation}
    r - r^c_{\rm gge} = \frac{3 m_{\rm gs} \pi}{8 \Lambda^3} m_{\rm gge} + \gamma_\Lambda \  m_{\rm gge}^2 
\end{equation}
where $\Lambda = \pi/a$ is the ultraviolet cutoff and $\gamma_\Lambda \approx 1$. 

This result shows that the GGE critical point coincides with the thermal one, giving $\nu = 1$ in $d = 3$. However, the critical scaling window in which such exponents can be observed is restricted by the condition $m_{\rm gge} \ll 3 m_{\rm gs} a^3 / (8 \pi^2)$ which becomes increasingly narrow for smaller lattice spacings $a$. Consequently, as illustrated in Fig.\,\ref{fig: Fig_SM_m2_phi_exponents}, for a small but finite quench depth $\delta r = |r-r_\mathrm{gge}^c|$, decreasing $a$ induces a crossover toward the mean-field exponents $\nu = 1/2$ and $\beta = 1/4$. This accounts for the numerical scaling behavior reported in Refs.~\cite{sciolla2013quantum, weidinger2017dynamical}.

\begin{figure*}
    \centering    \includegraphics[width=\linewidth]{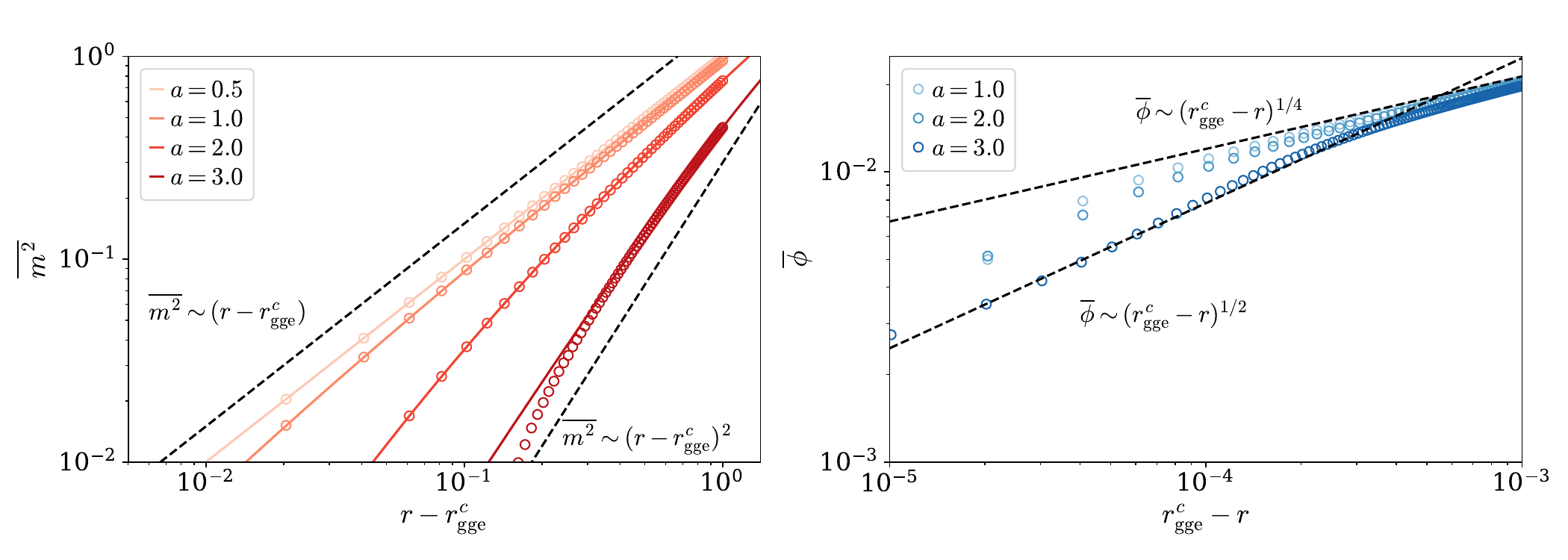}
    \caption{Panels (a) and (b) show, respectively, the scaling of the long-time average of the squared mass $m^2$ and of the order parameter $\phi = \eta_0 / \sqrt{N}$ as functions of $\delta r = |r - r_\mathrm{gge}^c|$ close to the dynamical transition (log–log scale), for different values of the lattice spacing $a$. The numerical results for $m^2$ obtained from the equations of motion (red squares) are in excellent agreement with the GGE prediction (solid lines). The system size is $ L = 5a\times 10^3$, $\lambda = 1$. As discussed in the main text, the scaling behavior consistent with the spherical-model universality class in $d=3$, $\overline{m^2}\sim (r-r_\mathrm{gge}^c)^{2}$, $\overline{\phi}\sim (r_\mathrm{gge}^c-r)^{1/2}$ emerges within the range of numerically accessible $\delta r$ only for sufficiently large lattice spacings ($a > 1$). Conversely, for smaller lattice spacings ($a < 1$), the numerical data display an effective scaling $\overline{m^2}\sim (r-r_\mathrm{gge}^c)$, $\overline{\phi}\sim (r_\mathrm{gge}^c-r)^{1/4}$ in agreement with previous findings in Refs. \cite{sciolla2013quantum,weidinger2017dynamical}.}
   \label{fig: Fig_SM_m2_phi_exponents}
\end{figure*}

\section{Spectrum in the QFT limit}
\noindent
We will now expose in details the argument sketched in the main text to compute the spectrum of the model in the quantum-field-theory (QFT) limit. In this limit, as the spectrum becomes continuous, that $u_l^{+}-u_l^- \sim L^{-1}$ for $l=0, \cdots,\mathcal{N}-2$, while the ultraviolet cutoff $\Lambda \sim a^{-1} \rightarrow \infty$ . 

In order to find the frequencies we have to express $\mathcal{H} = \sum_\mathbf{k} \epsilon_\mathbf{k}$ in terms of the action variables $J_l$. As explained in the main text, we introduce the auxiliary function 
\begin{equation}
    f(u) = \sqrt{(u+\mu) V(u)} \, . 
\end{equation}
This has $\mathcal{N}$ first-order poles for $u= \omega_l^2$, and $\mathcal{N}$ branch-cuts for $u_i^- < u < u_l^+$. If we consider the contour at infinity $C_\infty$ and we deform it on the real axis we have 
\begin{equation}
    \oint_{C_\infty} dz \ f(z) = \sum_l 2 \pi \text{i} \ \text{Res}_{\omega^2_l} f(z) + \int_{C_l} dz \ f(z) 
\end{equation}
where the $C_l$ are counter-clockwise contours which goes around the the $q_l^- < u < q_l^+$ branch-cut (see also Fig.1 in the main text). It is crucial that, in the QFT limit, the factor $\sqrt{\mu + u}$ can be considered constant along each branch cut. While for $l \neq \mathcal{N} -1$ this is simply a consequence of the fact that the spectral gap closes; for $l = \mathcal{N} -1$, the integration interval is located beyond the edge of the spectrum $\Lambda$ and thus unbounded. Yet our assumption becomes true in the field-theoretical limit in which $\Lambda \rightarrow \infty$: indeed, as $u_{\mathcal{N}-1}^{\pm} =O( \Lambda)$ one has $\sqrt{u+\mu} = \sqrt{\Lambda^2+\mu} + O(\Lambda^{-1})$. We have
\begin{equation}
    \int_{C_l} dz \ f(z) = - 2 \text{i} \int_{u_l^-}^{u_i^+} |f(u)| du \approx - 2 i \sqrt{\omega_l^2+\mu} \int_{u_l^-}^{u_l^+} \sqrt{-V(u)} du = -\frac{2 \pi \text{i} \lambda}{N} J_l \sqrt{\omega_l^2 + \mu}
\end{equation}
while 
\begin{equation}
    \text{Res}_{\omega^2_l} f(z) = - \frac{\lambda}{2N} \ell_l \sqrt{\omega_l^2 + \mu}
\end{equation}
on the other hand, since $f(z) \sim z + (r+\mu) - [\lambda \mathcal{H}/(2N)+ (\mu-r)^2/8] z^{-1} + O(z^{-2})$ one has 
\begin{equation}
   \oint_{C_\infty} dz \ f(z) = - 2 \pi \text{i} \left( \frac{\lambda}{2N} \mathcal{H} + \frac{(\mu-r)^2}{8} \right) 
\end{equation}
so that finally we get an expression for $\mathcal{H}$ in terms of the $J_l$,$\ell_l$ and $\mu$, namely 
\begin{equation} \label{eq:HJ_k}
    \mathcal{H} = - N \frac{(\mu-r)^2}{4 \lambda} + \sum_l \sqrt{\omega_l^2+\mu} \ (2J_l + \ell_l) \, .
\end{equation}
We have still to express the zero $\mu$ in terms of the $J_l$. To do so we consider the function 
\begin{equation}
    g(z) = \sqrt{\frac{V(z)}{z+\mu}} \, . 
\end{equation}
which has the exact branch and poles structures of $f(z)$, but a different behavior at infinity, namely $g(z) = 1 - (\mu - r)/2 z^{-1} + O(z^{-2})$. If we integrate it on the same contour thus we get
\begin{equation} \label{eq:mJk}
    \mu-r = \frac{\lambda}{N} \sum_l \frac{2 J_l + \ell_l}{\sqrt{\omega_l^2+\mu}} \, ,
\end{equation}
which is an implicit equation for $\mu$. Putting together Eq.\,\eqref{eq:HJ_k} and Eq.\,\eqref{eq:mJk} one has the form of the Hamiltonian $\mathcal{H}(J_l)$ in terms of the action variables $J_l, \ell_l$. Since the Hamiltonian depends only on $2 J_l + \ell_l$ we have a $\Omega_{l,2} = \Omega_{l,1}$; i.e. a set of doubly degenerate frequencies. The latter are given by
\begin{equation}
    \Omega_{l,1}  = \frac{\partial \mathcal{H}}{\partial \ell_k} = \sqrt{\omega_l^2 + \mu} + \frac{\partial \mu}{\partial \ell_l} \left( - \frac{\mu-r}{2 \lambda} + \frac{1}{2\mathcal{N}} \sum_l \frac{2J_l+\ell_l}{\sqrt{\omega_l^2+\mu}}  \right) = \sqrt{\omega_l^2 + \mu} \, . 
\end{equation}
Notice that we can interpret $\mu$, as the asymptotic value of the (squared) effective mass. On the other hand, by exploiting the definition of $\mu$, ($V(-\mu)=0$) one gets the self consistency equation in terms of the $\epsilon_l$, namely 
\begin{equation}
        0 = - \mu + r + \frac{\lambda}{N} \sum_\mathbf{k} \frac{\epsilon_\mathbf{k}}{\mu + \omega_\mathbf{k}^2} + \frac{\lambda^2}{4N^2} \sum_\mathbf{k} \frac{\ell_k^2}{(\mu+ \omega_\mathbf{k}^2)^2} \, .
\end{equation}
Unless $\mu \sim N^{-1}$, we see that that the last term on the right-hand-side is negligible for $N \rightarrow \infty$  so that one has 
\begin{equation}
        \bar{m}^2 = r + \frac{\lambda}{N} \sum_\mathbf{k} \frac{\epsilon_\mathbf{k}}{\bar{m}^2 + \omega_\mathbf{k}^2} \, . 
\end{equation}
from which $\mu = m^2_{\rm gge}$.

\section*{Lattice case}
\noindent
In this section we will derive the form assumed by the potential $V(u)$ (Eq.\,(9) of the main text) 
\begin{equation} \label{eq:SMVu}
    V(u) = u + r - \frac{\lambda}{N} \sum_{\mathbf{k}} \frac{\epsilon_\mathbf{k}}{u-\omega^2_\mathbf{k}} + \frac{\lambda^2}{4N^2}\sum_\mathbf{k} \frac{\ell_\mathbf{k}^2}{(u-\omega^2_\mathbf{k})^2}  .
\end{equation}
in the thermodynamic limit. We start by noticing that far from the singularities ($u\notin [0,\Lambda]$), we can replace the sum with integrals, obtaining
\begin{equation}
    V_{\rm reg}(u) = u + r +\lambda \int_0^\Lambda \frac{dE \  \rho(E) \epsilon(E)}{u-E} ,
\end{equation}
$\rho(u)$ being the density of states of the phonon dispersion $\omega_l^2$. Notice that $V_{\rm reg} (z)$ is analytic evweerywhere but in the interval $z \in [0,\Lambda]$, where it has a branch cut. Moreover for $u \in \mathbb{R} / [0,\Lambda]$ we inherit he polynomial structure $V(u) < 0 $ for $u < -\mu$ and $u^-_{\mathcal{N}-1}< u < u_{\mathcal{N}-1}^+$. Let us notice however, that in this limit $u^-_{\mathcal{N}-1} \rightarrow \Lambda$. 

As the interval $u \in [\omega_l^2, \omega_{l+1}^2]$ is $O(N^{-1})$, in the $N \rightarrow \infty$ limit we have to take into account the singular structure of the potential. It is thus convenient to parametrize 
\begin{equation}
    u_l = \omega_l^2 + \frac{\Delta \omega^2_l}{\pi}  \theta_l
\end{equation}
with $\Delta \omega^2_{l} = \omega_{l+1}^2 - \omega_l^2$ and $\theta_l \in [0, \pi]$. Let us notice that $\Delta \omega^2_{l} = O (N^{-1})$, and, more exactly, $(N \Delta \omega_l^2)^ {-1} = \rho(u)$, i.e. the density of states. 
In this case, in the limit $N \rightarrow \infty$ we have (see below)
\begin{equation} \label{eq:Vtheta}
    V_l (\theta) = \omega_l^2 + r - \lambda \ \text{p.v.} \int  dE \ \frac{\rho(E) \epsilon(E)}{E-\omega_l^2}   -  \frac{\lambda \pi \rho(\omega_l^2) \epsilon_l}{\tan \theta}  + \frac{\left(\lambda \pi \rho(\omega_l^2) \ell_l \right)^2}{4\sin^2 \theta } \, .
\end{equation}

\textit{Proof of Eq.\eqref{eq:Vtheta}:} First we notice that the linear term $2(u + r)$ in Eq.\,\eqref{eq:SMVu} can be approximated with a constant $2(\omega_l^2 + r)$. Instead, the second term in Eq.\,\eqref{eq:SMVu} 
\begin{equation}
     \frac{1}{N} \sum_{l'} \frac{\epsilon_{l'}}{u_l - \omega_{l'}^2}
\end{equation}
can be separated into a singular part, containing the poles, and a non-singular, slowly-varying, one, simply given by
\begin{equation}
       \text{p.v.} \int \frac{d^d \mathbf{k}}{(2 \pi)^d} \frac{\epsilon_\mathbf{k'}}{\omega^2_\mathbf{k} - \omega^2_\mathbf{k'}} =  \text{p.v.} \int  du \ \frac{\rho(u') \epsilon(u')}{u-u'} \equiv \mathbb{H}_u (\rho \epsilon)
\end{equation}
(notice that this also can be approximated as a constant in the interval considered). The singular part, coming from the poles in the vicinity of $u= \omega_l^2$, and can be estimated as follows
\begin{equation}
    \frac{1}{N} \sum_{l'} \frac{\epsilon_{l'}}{u_l - \omega_{l'}^2} =  \sum_{l'} \frac{\epsilon_{l'}}{N(\omega_l^2 - \omega_{l'}^2) + N \Delta \omega^2_l \theta/ \pi} \, . 
\end{equation}
In the large $N$ limit the $\omega_l$ and $\epsilon_l$ varies slowly with $l$; on the other hand the contribution of the terms with $l^\prime \gg l$ vanishes, so that one can approximate $\epsilon_{l'}$ $N(\omega^2_{l} - \omega_{l'}^2) \sim -N \Delta \omega^2_l (l-l')$. We have that 
\begin{equation}
\begin{split}
    \frac{1}{N} \sum_{l'} \frac{\epsilon_{l'}}{u_l - \omega_{l'}^2} =  \frac{\epsilon_{l}}{N \Delta \omega^2_l} \sum_{l'} \frac{1}{l - l'  + \theta/\pi}  =   \pi \rho(\omega_l^2) \epsilon_l \cot \theta
    \end{split}
\end{equation}
By deriving both sides with respect to $\theta$, and replacing $\epsilon_l$ with $\ell_l$ one finds 
\begin{equation}
\begin{split}
    \frac{1}{N^2} \sum_{l'} \frac{\ell^2_{l'}}{(u_l - \omega_{l'}^2)^2}
    = \pi^2 \rho^2(\omega_l^2) \frac{\ell^2_l}{\sin^2 \theta} \, 
\end{split}
\end{equation}
which is the last term of Eq.\,\eqref{eq:SMVu}.

\subsection{Action variables}
We have that 
\begin{equation}
    J_{\mathcal{N}-1} = \frac{N}{\pi \lambda} \int_{\Lambda}^{b} du \ \sqrt{-V_{\rm reg}(u)} 
\end{equation}
(where, as before, $ b= u^+_{\mathcal{N}-1}$, the largest zero of $V_{\rm reg}(u)$). For $l=0, \dots,  \mathcal{N}-2$, it is convenient to write the integral in $\theta$
\begin{equation}
\begin{split}
     J_l &= \frac{N\Delta \omega_l^2}{\pi^2 \lambda} \int^{\theta^{+}}_{\theta^{-}} d \theta \sqrt{- V_l (\theta)} \\
        &=  - \frac{\ell_l}{2}   +  \frac{1}{\pi \lambda \rho(\omega_l^2)}  \Im \sqrt{\omega_l^2+r - \lambda \ \text{p.v.} \int  dE' \ \frac{\rho(E') \epsilon(E')}{E'-\omega_l^2}  + i \lambda \pi \rho(\omega_l^2) \epsilon_l} \\
     &= - \frac{\ell_l}{2}  + \frac{1}{\pi \lambda \rho(\omega_l^2)} \Im \sqrt{\omega_l^2+r - \lambda \int \frac{dE'}{E^\prime-\omega_l^2 - i 0^+} \rho(E^\prime) \epsilon(E^\prime)}
\end{split}
\end{equation}
or,
\begin{equation}
    \pi \lambda \rho(\omega_l^2) \left( J_\ell + \frac{1}{2}\ell_l \right) =\Im \sqrt{V_{\rm reg} (\omega_l^2 + i 0)} \, . 
\end{equation}
In the continuum limit, we introduce again $E = \omega_l^2 \in [0,\Lambda]$ and $\mathcal{J}(E) = (J_l + \ell_l/2)$
\begin{equation}
    \pi \lambda \rho(E) \mathcal{J}(E) = \Im \sqrt{V_{\rm reg} (E+ i 0)} \, . 
\end{equation}
By carefully handling the singularity structure of $\sqrt{V_{\rm reg} (z)}$, this expression can be inverted with the use of the Cauchy–Plemelj formula. In particular, we have that, outside the interval $[0,\Lambda]$, $\sqrt{V(u)}$ has branch cuts for $u < -\mu$ and $u \in [\Lambda,b]$, while at $z= \infty$ $\sqrt{V(z)} \sim \sqrt{z}$. We have thus 
\begin{equation}
    V(z) = \frac{(z + \mu) (z-\Lambda)}{z-b} \left( 1 + \lambda \int_0^\Lambda  \frac{dE}{\sqrt{E+\mu}} \sqrt{\frac{b-E} {\Lambda-E}} \frac{\rho(E) \mathcal{J}(E)}{E-z} \right)^2 
\end{equation}
Which can alternatively be written by coming back to the discrete formulation as 
\begin{equation}
    V(z) = \frac{(z + \mu) (z-\Lambda)}{z-b} \left( 1 + \frac{\lambda}{2N} \sum_{l =0}^{\mathcal{N}-2} \frac{2 J_l + \ell_l}{\omega_l^2-z}  \sqrt{\frac{b-\omega_l^2} {(\Lambda-\omega_l^2)(\mu+\omega_l^2)}}   \right)^2 \, . 
\end{equation} 
By expanding both sides around $z = \infty$, and remembering that $V(z) = z + r - \lambda \mathcal{H}/N z^{-1} + O(z^{-2})$, we find 
\begin{equation} \label{eq:mJklattice}
    \mu - r + b - \Lambda  = \frac{\lambda}{N} \sum_{l =0}^{\mathcal{N}-2} \frac{2 J_l + \ell_l}{\sqrt{\omega_l^2+\mu}} \sqrt{\frac{b-\omega_l^2} {\Lambda-\omega_l^2}} \, ,
\end{equation}
and
\begin{equation} \label{eq:HJ_klattice}
    \mathcal{H} =  -\frac{N}{ \lambda} (\Lambda+r)(b-\Lambda)-  \frac{N}{4 \lambda} (\mu-r +b-\Lambda)^2+ \sum_{l =0}^{\mathcal{N}-2} \sqrt{\omega_l^2+\mu} \ \sqrt{\frac{b-\omega_l^2} {\Lambda-\omega_l^2}}\ (2J_l + \ell_l) ;
\end{equation}
which are the lattice analogous of the relations\,\eqref{eq:mJk} and \eqref{eq:HJ_k} respectively. 

Yet, since  on the lattice we still have the free parameter $b$, we ought express this in terms of the $J_k$. This can be done by exploiting the last action variable $J_{\mathcal{N}-1}$
\begin{equation} \label{eq:JN-1}
    J_{\mathcal{N}-1} = \frac{N}{\pi \lambda} \int_\Lambda^b du \ \sqrt{\frac{(\mu + u)(\Lambda-u)}{u-b}} + \frac{1}{2 \pi} \sum_{l=0}^{\mathcal{N}-2}  (2 J_l + \ell_l) \sqrt{\frac{b-\omega_l^2} {(\Lambda-\omega_l^2)(\mu+\omega_l^2)}} \int_\Lambda^b \frac{du}{\omega_l^2-u} \ \sqrt{\frac{(\mu + u)(\Lambda-u)}{u-b}} \, .
 \end{equation}
Eq.\,\eqref{eq:HJ_klattice}, together with the conditions \eqref{eq:mJklattice} and \eqref{eq:JN-1}, allow to compute directly the isolated frequency $\bar{\Omega} = \partial \mathcal{H}/\partial J_{\mathcal{N}-1}$  and the rest of the spectrum $\Omega_{l,1} = \partial \mathcal{H}/\partial J_l$,$\Omega_{l,2} = \partial \mathcal{H}/\partial \ell_l$.
 
\section{Numerical methods}
The numerical results shown in the main text are obtained by numerically integrating the equations of motion in Eqs.\,(4) of the main text, which govern the evolution of the complex variables $\eta_k$ and $p_k$  , using a fourth-order Runge–Kutta algorithm with suitably small time steps. 

In order to access larger system sizes in three dimensions ($d = 3$), we employ the Debye approximation,\cite{debye1912zurtheorie,ashcroft1976}, which consists in replacing the cubic Brillouin zone spanned by $\mathbf{k} = (k_x, k_y, k_z)$ with a sphere of radius $\Lambda$. This approximation preserves the low-momentum density of states while significantly reducing the computational cost. Within this scheme, the problem is effectively reduced to $2L/a$ degrees of freedom, corresponding to the complex variables $\eta_k$ and $p_k$, labeled by 
\begin{align}
    k =  0,\frac{\pi}{L},\frac{2\pi}{L}\dots,\frac{\pi}{a} \equiv \Lambda,
\end{align}
which evolve according to
\begin{align}
    p_k(t) = \dot{\eta}_k(t),\quad\dot{p}_k(t) = -\left(m^2(t)+\omega_k^2\right)\eta_k(t),
\end{align}
with the squared mass $m^2(t)$ computed self-consistently at each time step through
\begin{align}
    m^2(t) = r+\frac{a d}{L}\sum_k \left(\frac{ak}{\pi}\right)^{(d - 1)}|\eta_k(t)|^2 \, .
\end{align}
The order parameter corresponds to the normalized zero mode, $\phi = \sqrt{a/L}\eta_0$.

The initial conditions are fixed by the ground state conditions, while for the zero mode $k = 0$
\begin{equation}
    \eta_{k=0}(0) = \sqrt{\frac{\lambda}{2}}\left(m^2_\mathrm{gs}(r_{-}) + \omega_k^2\right)^{-1/4},\hspace{0.5cm}
    p_{k=0}(0) = 0
\end{equation}
With these choices, $\eta_{0}(t)$ and $p_{0}(t)$ remain real-valued at all times.




\begin{thebibliography}{69}%
\makeatletter
\providecommand \@ifxundefined [1]{%
 \@ifx{#1\undefined}
}%
\providecommand \@ifnum [1]{%
 \ifnum #1\expandafter \@firstoftwo
 \else \expandafter \@secondoftwo
 \fi
}%
\providecommand \@ifx [1]{%
 \ifx #1\expandafter \@firstoftwo
 \else \expandafter \@secondoftwo
 \fi
}%
\providecommand \natexlab [1]{#1}%
\providecommand \enquote  [1]{``#1''}%
\providecommand \bibnamefont  [1]{#1}%
\providecommand \bibfnamefont [1]{#1}%
\providecommand \citenamefont [1]{#1}%
\providecommand \href@noop [0]{\@secondoftwo}%
\providecommand \href [0]{\begingroup \@sanitize@url \@href}%
\providecommand \@href[1]{\@@startlink{#1}\@@href}%
\providecommand \@@href[1]{\endgroup#1\@@endlink}%
\providecommand \@sanitize@url [0]{\catcode `\\12\catcode `\$12\catcode `\&12\catcode `\#12\catcode `\^12\catcode `\_12\catcode `\%12\relax}%
\providecommand \@@startlink[1]{}%
\providecommand \@@endlink[0]{}%
\providecommand \url  [0]{\begingroup\@sanitize@url \@url }%
\providecommand \@url [1]{\endgroup\@href {#1}{\urlprefix }}%
\providecommand \urlprefix  [0]{URL }%
\providecommand \Eprint [0]{\href }%
\providecommand \doibase [0]{https://doi.org/}%
\providecommand \selectlanguage [0]{\@gobble}%
\providecommand \bibinfo  [0]{\@secondoftwo}%
\providecommand \bibfield  [0]{\@secondoftwo}%
\providecommand \translation [1]{[#1]}%
\providecommand \BibitemOpen [0]{}%
\providecommand \bibitemStop [0]{}%
\providecommand \bibitemNoStop [0]{.\EOS\space}%
\providecommand \EOS [0]{\spacefactor3000\relax}%
\providecommand \BibitemShut  [1]{\csname bibitem#1\endcsname}%
\let\auto@bib@innerbib\@empty
\bibitem [{\citenamefont {Kubo}(1991)}]{kubo1991nonequilibrium}%
  \BibitemOpen
  \bibfield  {author} {\bibinfo {author} {\bibfnamefont {R.}~\bibnamefont {Kubo}},\ }\bibfield  {title} {\bibinfo {title} {Nonequili-brium statistical mechanics},\ }\href@noop {} {\bibfield  {journal} {\bibinfo  {journal} {Statistical Physics}\ }\textbf {\bibinfo {volume} {11}} (\bibinfo {year} {1991})}\BibitemShut {NoStop}%
\bibitem [{\citenamefont {Schuster}\ and\ \citenamefont {Just}(1988)}]{schuster1988deterministic}%
  \BibitemOpen
  \bibfield  {author} {\bibinfo {author} {\bibfnamefont {H.~G.}\ \bibnamefont {Schuster}}\ and\ \bibinfo {author} {\bibfnamefont {W.}~\bibnamefont {Just}},\ }\href@noop {} {\emph {\bibinfo {title} {Deterministic chaos: an introduction}}},\ Vol.~\bibinfo {volume} {2}\ (\bibinfo  {publisher} {Wiley Online Library},\ \bibinfo {year} {1988})\BibitemShut {NoStop}%
\bibitem [{\citenamefont {Gallavotti}(2007)}]{gallavotti2007fermi}%
  \BibitemOpen
  \bibfield  {author} {\bibinfo {author} {\bibfnamefont {G.}~\bibnamefont {Gallavotti}},\ }\href@noop {} {\emph {\bibinfo {title} {The Fermi-Pasta-Ulam problem: a status report}}},\ Vol.\ \bibinfo {volume} {728}\ (\bibinfo  {publisher} {Springer},\ \bibinfo {year} {2007})\BibitemShut {NoStop}%
\bibitem [{\citenamefont {von Neumann}(2010)}]{vonNeumann2010proof}%
  \BibitemOpen
  \bibfield  {author} {\bibinfo {author} {\bibfnamefont {J.}~\bibnamefont {von Neumann}},\ }\bibfield  {title} {\bibinfo {title} {Proof of the ergodic theorem and the h-theorem in quantum mechanics: Translation of: Beweis des ergodensatzes und des h-theorems in der neuen mechanik},\ }\href {https://doi.org/10.1140/epjh/e2010-00008-5} {\bibfield  {journal} {\bibinfo  {journal} {The European Physical Journal H}\ }\textbf {\bibinfo {volume} {35}},\ \bibinfo {pages} {201–237} (\bibinfo {year} {2010})}\BibitemShut {NoStop}%
\bibitem [{\citenamefont {Srednicki}(1994)}]{srednicki1994chaos}%
  \BibitemOpen
  \bibfield  {author} {\bibinfo {author} {\bibfnamefont {M.}~\bibnamefont {Srednicki}},\ }\bibfield  {title} {\bibinfo {title} {Chaos and quantum thermalization},\ }\href@noop {} {\bibfield  {journal} {\bibinfo  {journal} {Phys. Rev. E}\ }\textbf {\bibinfo {volume} {50}},\ \bibinfo {pages} {888} (\bibinfo {year} {1994})}\BibitemShut {NoStop}%
\bibitem [{\citenamefont {Rigol}\ and\ \citenamefont {Srednicki}(2012)}]{rigol2012alternatives}%
  \BibitemOpen
  \bibfield  {author} {\bibinfo {author} {\bibfnamefont {M.}~\bibnamefont {Rigol}}\ and\ \bibinfo {author} {\bibfnamefont {M.}~\bibnamefont {Srednicki}},\ }\bibfield  {title} {\bibinfo {title} {Alternatives to eigenstate thermalization},\ }\href {https://doi.org/10.1103/PhysRevLett.108.110601} {\bibfield  {journal} {\bibinfo  {journal} {Phys. Rev. Lett.}\ }\textbf {\bibinfo {volume} {108}},\ \bibinfo {pages} {110601} (\bibinfo {year} {2012})}\BibitemShut {NoStop}%
\bibitem [{\citenamefont {Goldstein}\ \emph {et~al.}(2006)\citenamefont {Goldstein}, \citenamefont {Lebowitz}, \citenamefont {Tumulka},\ and\ \citenamefont {Zangh\`{\i}}}]{goldstein2006canonical}%
  \BibitemOpen
  \bibfield  {author} {\bibinfo {author} {\bibfnamefont {S.}~\bibnamefont {Goldstein}}, \bibinfo {author} {\bibfnamefont {J.~L.}\ \bibnamefont {Lebowitz}}, \bibinfo {author} {\bibfnamefont {R.}~\bibnamefont {Tumulka}},\ and\ \bibinfo {author} {\bibfnamefont {N.}~\bibnamefont {Zangh\`{\i}}},\ }\bibfield  {title} {\bibinfo {title} {Canonical typicality},\ }\href {https://doi.org/10.1103/PhysRevLett.96.050403} {\bibfield  {journal} {\bibinfo  {journal} {Phys. Rev. Lett.}\ }\textbf {\bibinfo {volume} {96}},\ \bibinfo {pages} {050403} (\bibinfo {year} {2006})}\BibitemShut {NoStop}%
\bibitem [{\citenamefont {Manmana}\ \emph {et~al.}(2007)\citenamefont {Manmana}, \citenamefont {Wessel}, \citenamefont {Noack},\ and\ \citenamefont {Muramatsu}}]{manmana2007strongly}%
  \BibitemOpen
  \bibfield  {author} {\bibinfo {author} {\bibfnamefont {S.~R.}\ \bibnamefont {Manmana}}, \bibinfo {author} {\bibfnamefont {S.}~\bibnamefont {Wessel}}, \bibinfo {author} {\bibfnamefont {R.~M.}\ \bibnamefont {Noack}},\ and\ \bibinfo {author} {\bibfnamefont {A.}~\bibnamefont {Muramatsu}},\ }\bibfield  {title} {\bibinfo {title} {Strongly correlated fermions after a quantum quench},\ }\href {https://doi.org/10.1103/PhysRevLett.98.210405} {\bibfield  {journal} {\bibinfo  {journal} {Phys. Rev. Lett.}\ }\textbf {\bibinfo {volume} {98}},\ \bibinfo {pages} {210405} (\bibinfo {year} {2007})}\BibitemShut {NoStop}%
\bibitem [{\citenamefont {Moeckel}\ and\ \citenamefont {Kehrein}(2008)}]{moeckel2008interaction}%
  \BibitemOpen
  \bibfield  {author} {\bibinfo {author} {\bibfnamefont {M.}~\bibnamefont {Moeckel}}\ and\ \bibinfo {author} {\bibfnamefont {S.}~\bibnamefont {Kehrein}},\ }\bibfield  {title} {\bibinfo {title} {Interaction quench in the hubbard model},\ }\href {https://doi.org/10.1103/PhysRevLett.100.175702} {\bibfield  {journal} {\bibinfo  {journal} {Phys. Rev. Lett.}\ }\textbf {\bibinfo {volume} {100}},\ \bibinfo {pages} {175702} (\bibinfo {year} {2008})}\BibitemShut {NoStop}%
\bibitem [{\citenamefont {Biroli}\ \emph {et~al.}(2010)\citenamefont {Biroli}, \citenamefont {Kollath},\ and\ \citenamefont {L\"auchli}}]{biroli2010effect}%
  \BibitemOpen
  \bibfield  {author} {\bibinfo {author} {\bibfnamefont {G.}~\bibnamefont {Biroli}}, \bibinfo {author} {\bibfnamefont {C.}~\bibnamefont {Kollath}},\ and\ \bibinfo {author} {\bibfnamefont {A.~M.}\ \bibnamefont {L\"auchli}},\ }\bibfield  {title} {\bibinfo {title} {Effect of rare fluctuations on the thermalization of isolated quantum systems},\ }\href {https://doi.org/10.1103/PhysRevLett.105.250401} {\bibfield  {journal} {\bibinfo  {journal} {Phys. Rev. Lett.}\ }\textbf {\bibinfo {volume} {105}},\ \bibinfo {pages} {250401} (\bibinfo {year} {2010})}\BibitemShut {NoStop}%
\bibitem [{\citenamefont {Khatami}\ \emph {et~al.}(2013)\citenamefont {Khatami}, \citenamefont {Pupillo}, \citenamefont {Srednicki},\ and\ \citenamefont {Rigol}}]{khatami2013fluctuation}%
  \BibitemOpen
  \bibfield  {author} {\bibinfo {author} {\bibfnamefont {E.}~\bibnamefont {Khatami}}, \bibinfo {author} {\bibfnamefont {G.}~\bibnamefont {Pupillo}}, \bibinfo {author} {\bibfnamefont {M.}~\bibnamefont {Srednicki}},\ and\ \bibinfo {author} {\bibfnamefont {M.}~\bibnamefont {Rigol}},\ }\bibfield  {title} {\bibinfo {title} {Fluctuation-dissipation theorem in an isolated system of quantum dipolar bosons after a quench},\ }\href {https://doi.org/10.1103/PhysRevLett.111.050403} {\bibfield  {journal} {\bibinfo  {journal} {Phys. Rev. Lett.}\ }\textbf {\bibinfo {volume} {111}},\ \bibinfo {pages} {050403} (\bibinfo {year} {2013})}\BibitemShut {NoStop}%
\bibitem [{\citenamefont {Yoshizawa}\ \emph {et~al.}(2018)\citenamefont {Yoshizawa}, \citenamefont {Iyoda},\ and\ \citenamefont {Sagawa}}]{yoshizawa2018numerical}%
  \BibitemOpen
  \bibfield  {author} {\bibinfo {author} {\bibfnamefont {T.}~\bibnamefont {Yoshizawa}}, \bibinfo {author} {\bibfnamefont {E.}~\bibnamefont {Iyoda}},\ and\ \bibinfo {author} {\bibfnamefont {T.}~\bibnamefont {Sagawa}},\ }\bibfield  {title} {\bibinfo {title} {Numerical large deviation analysis of the eigenstate thermalization hypothesis},\ }\href {https://doi.org/10.1103/PhysRevLett.120.200604} {\bibfield  {journal} {\bibinfo  {journal} {Phys. Rev. Lett.}\ }\textbf {\bibinfo {volume} {120}},\ \bibinfo {pages} {200604} (\bibinfo {year} {2018})}\BibitemShut {NoStop}%
\bibitem [{\citenamefont {Foini}\ and\ \citenamefont {Kurchan}(2019)}]{foini2019eigenstate}%
  \BibitemOpen
  \bibfield  {author} {\bibinfo {author} {\bibfnamefont {L.}~\bibnamefont {Foini}}\ and\ \bibinfo {author} {\bibfnamefont {J.}~\bibnamefont {Kurchan}},\ }\bibfield  {title} {\bibinfo {title} {Eigenstate thermalization and rotational invariance in ergodic quantum systems},\ }\href {https://doi.org/10.1103/PhysRevLett.123.260601} {\bibfield  {journal} {\bibinfo  {journal} {Phys. Rev. Lett.}\ }\textbf {\bibinfo {volume} {123}},\ \bibinfo {pages} {260601} (\bibinfo {year} {2019})}\BibitemShut {NoStop}%
\bibitem [{\citenamefont {Pappalardi}\ \emph {et~al.}(2022)\citenamefont {Pappalardi}, \citenamefont {Foini},\ and\ \citenamefont {Kurchan}}]{pappalardi2022eigenstate}%
  \BibitemOpen
  \bibfield  {author} {\bibinfo {author} {\bibfnamefont {S.}~\bibnamefont {Pappalardi}}, \bibinfo {author} {\bibfnamefont {L.}~\bibnamefont {Foini}},\ and\ \bibinfo {author} {\bibfnamefont {J.}~\bibnamefont {Kurchan}},\ }\bibfield  {title} {\bibinfo {title} {Eigenstate thermalization hypothesis and free probability},\ }\href {https://doi.org/10.1103/PhysRevLett.129.170603} {\bibfield  {journal} {\bibinfo  {journal} {Phys. Rev. Lett.}\ }\textbf {\bibinfo {volume} {129}},\ \bibinfo {pages} {170603} (\bibinfo {year} {2022})}\BibitemShut {NoStop}%
\bibitem [{\citenamefont {Pappalardi}\ \emph {et~al.}(2025)\citenamefont {Pappalardi}, \citenamefont {Fritzsch},\ and\ \citenamefont {Prosen}}]{pappalardi2025full}%
  \BibitemOpen
  \bibfield  {author} {\bibinfo {author} {\bibfnamefont {S.}~\bibnamefont {Pappalardi}}, \bibinfo {author} {\bibfnamefont {F.}~\bibnamefont {Fritzsch}},\ and\ \bibinfo {author} {\bibfnamefont {T.~c.~v.}\ \bibnamefont {Prosen}},\ }\bibfield  {title} {\bibinfo {title} {Full eigenstate thermalization via free cumulants in quantum lattice systems},\ }\href {https://doi.org/10.1103/PhysRevLett.134.140404} {\bibfield  {journal} {\bibinfo  {journal} {Phys. Rev. Lett.}\ }\textbf {\bibinfo {volume} {134}},\ \bibinfo {pages} {140404} (\bibinfo {year} {2025})}\BibitemShut {NoStop}%
\bibitem [{\citenamefont {Mitra}(2018)}]{mitra2018quantum}%
  \BibitemOpen
  \bibfield  {author} {\bibinfo {author} {\bibfnamefont {A.}~\bibnamefont {Mitra}},\ }\bibfield  {title} {\bibinfo {title} {Quantum quench dynamics},\ }\href {https://doi.org/10.1146/annurev-conmatphys-031016-025451} {\bibfield  {journal} {\bibinfo  {journal} {Annual Review of Condensed Matter Physics}\ }\textbf {\bibinfo {volume} {9}},\ \bibinfo {pages} {245–259} (\bibinfo {year} {2018})}\BibitemShut {NoStop}%
\bibitem [{\citenamefont {Polkovnikov}\ \emph {et~al.}(2011)\citenamefont {Polkovnikov}, \citenamefont {Sengupta}, \citenamefont {Silva},\ and\ \citenamefont {Vengalattore}}]{polkovnikov2011colloquium}%
  \BibitemOpen
  \bibfield  {author} {\bibinfo {author} {\bibfnamefont {A.}~\bibnamefont {Polkovnikov}}, \bibinfo {author} {\bibfnamefont {K.}~\bibnamefont {Sengupta}}, \bibinfo {author} {\bibfnamefont {A.}~\bibnamefont {Silva}},\ and\ \bibinfo {author} {\bibfnamefont {M.}~\bibnamefont {Vengalattore}},\ }\bibfield  {title} {\bibinfo {title} {Colloquium: Nonequilibrium dynamics of closed interacting quantum systems},\ }\href {https://doi.org/10.1103/RevModPhys.83.863} {\bibfield  {journal} {\bibinfo  {journal} {Rev. Mod. Phys.}\ }\textbf {\bibinfo {volume} {83}},\ \bibinfo {pages} {863} (\bibinfo {year} {2011})}\BibitemShut {NoStop}%
\bibitem [{\citenamefont {Calabrese}\ and\ \citenamefont {Cardy}(2006)}]{calabrese2006time}%
  \BibitemOpen
  \bibfield  {author} {\bibinfo {author} {\bibfnamefont {P.}~\bibnamefont {Calabrese}}\ and\ \bibinfo {author} {\bibfnamefont {J.}~\bibnamefont {Cardy}},\ }\bibfield  {title} {\bibinfo {title} {Time dependence of correlation functions following a quantum quench},\ }\href {https://doi.org/10.1103/PhysRevLett.96.136801} {\bibfield  {journal} {\bibinfo  {journal} {Phys. Rev. Lett.}\ }\textbf {\bibinfo {volume} {96}},\ \bibinfo {pages} {136801} (\bibinfo {year} {2006})}\BibitemShut {NoStop}%
\bibitem [{\citenamefont {Calabrese}\ and\ \citenamefont {Cardy}(2007)}]{calabrese2007quantum}%
  \BibitemOpen
  \bibfield  {author} {\bibinfo {author} {\bibfnamefont {P.}~\bibnamefont {Calabrese}}\ and\ \bibinfo {author} {\bibfnamefont {J.}~\bibnamefont {Cardy}},\ }\bibfield  {title} {\bibinfo {title} {Quantum quenches in extended systems},\ }\href {https://doi.org/10.1088/1742-5468/2007/06/p06008} {\bibfield  {journal} {\bibinfo  {journal} {Journal of Statistical Mechanics: Theory and Experiment}\ }\textbf {\bibinfo {volume} {2007}},\ \bibinfo {pages} {P06008–P06008} (\bibinfo {year} {2007})}\BibitemShut {NoStop}%
\bibitem [{\citenamefont {Gambassi}\ and\ \citenamefont {Silva}(2012)}]{gambassi2012large}%
  \BibitemOpen
  \bibfield  {author} {\bibinfo {author} {\bibfnamefont {A.}~\bibnamefont {Gambassi}}\ and\ \bibinfo {author} {\bibfnamefont {A.}~\bibnamefont {Silva}},\ }\bibfield  {title} {\bibinfo {title} {Large deviations and universality in quantum quenches},\ }\href {https://doi.org/10.1103/PhysRevLett.109.250602} {\bibfield  {journal} {\bibinfo  {journal} {Phys. Rev. Lett.}\ }\textbf {\bibinfo {volume} {109}},\ \bibinfo {pages} {250602} (\bibinfo {year} {2012})}\BibitemShut {NoStop}%
\bibitem [{\citenamefont {Solfanelli}\ and\ \citenamefont {Defenu}(2025)}]{solfanelli2025universal}%
  \BibitemOpen
  \bibfield  {author} {\bibinfo {author} {\bibfnamefont {A.}~\bibnamefont {Solfanelli}}\ and\ \bibinfo {author} {\bibfnamefont {N.}~\bibnamefont {Defenu}},\ }\bibfield  {title} {\bibinfo {title} {Universal work statistics in long-range interacting quantum systems},\ }\href {https://doi.org/10.1103/PhysRevLett.134.030402} {\bibfield  {journal} {\bibinfo  {journal} {Phys. Rev. Lett.}\ }\textbf {\bibinfo {volume} {134}},\ \bibinfo {pages} {030402} (\bibinfo {year} {2025})}\BibitemShut {NoStop}%
\bibitem [{\citenamefont {Gambassi}\ and\ \citenamefont {Calabrese}(2011)}]{gambassi2011quantum}%
  \BibitemOpen
  \bibfield  {author} {\bibinfo {author} {\bibfnamefont {A.}~\bibnamefont {Gambassi}}\ and\ \bibinfo {author} {\bibfnamefont {P.}~\bibnamefont {Calabrese}},\ }\bibfield  {title} {\bibinfo {title} {Quantum quenches as classical critical films},\ }\href {https://doi.org/10.1209/0295-5075/95/66007} {\bibfield  {journal} {\bibinfo  {journal} {Europhysics Letters}\ }\textbf {\bibinfo {volume} {95}},\ \bibinfo {pages} {66007} (\bibinfo {year} {2011})}\BibitemShut {NoStop}%
\bibitem [{\citenamefont {Jona-Lasinio}\ and\ \citenamefont {Presilla}(1996)}]{lasinio1996chaotic}%
  \BibitemOpen
  \bibfield  {author} {\bibinfo {author} {\bibfnamefont {G.}~\bibnamefont {Jona-Lasinio}}\ and\ \bibinfo {author} {\bibfnamefont {C.}~\bibnamefont {Presilla}},\ }\bibfield  {title} {\bibinfo {title} {Chaotic properties of quantum many-body systems in the thermodynamic limit},\ }\href {https://doi.org/10.1103/PhysRevLett.77.4322} {\bibfield  {journal} {\bibinfo  {journal} {Phys. Rev. Lett.}\ }\textbf {\bibinfo {volume} {77}},\ \bibinfo {pages} {4322} (\bibinfo {year} {1996})}\BibitemShut {NoStop}%
\bibitem [{\citenamefont {Lenci}(1996)}]{lenci1996ergodic}%
  \BibitemOpen
  \bibfield  {author} {\bibinfo {author} {\bibfnamefont {M.}~\bibnamefont {Lenci}},\ }\bibfield  {title} {\bibinfo {title} {Ergodic properties of the quantum ideal gas in the maxwell–boltzmann statistics},\ }\href {https://doi.org/10.1063/1.531684} {\bibfield  {journal} {\bibinfo  {journal} {Journal of Mathematical Physics}\ }\textbf {\bibinfo {volume} {37}},\ \bibinfo {pages} {5136–5157} (\bibinfo {year} {1996})}\BibitemShut {NoStop}%
\bibitem [{\citenamefont {Graffi}\ and\ \citenamefont {Martinez}(1996)}]{graffi1996ergodic}%
  \BibitemOpen
  \bibfield  {author} {\bibinfo {author} {\bibfnamefont {S.}~\bibnamefont {Graffi}}\ and\ \bibinfo {author} {\bibfnamefont {A.}~\bibnamefont {Martinez}},\ }\bibfield  {title} {\bibinfo {title} {Ergodic properties of infinite quantum harmonic crystals: An analytic approach},\ }\href {https://doi.org/10.1063/1.531741} {\bibfield  {journal} {\bibinfo  {journal} {Journal of Mathematical Physics}\ }\textbf {\bibinfo {volume} {37}},\ \bibinfo {pages} {5111–5135} (\bibinfo {year} {1996})}\BibitemShut {NoStop}%
\bibitem [{\citenamefont {Rigol}\ \emph {et~al.}(2007)\citenamefont {Rigol}, \citenamefont {Dunjko}, \citenamefont {Yurovsky},\ and\ \citenamefont {Olshanii}}]{rigol2007relaxation}%
  \BibitemOpen
  \bibfield  {author} {\bibinfo {author} {\bibfnamefont {M.}~\bibnamefont {Rigol}}, \bibinfo {author} {\bibfnamefont {V.}~\bibnamefont {Dunjko}}, \bibinfo {author} {\bibfnamefont {V.}~\bibnamefont {Yurovsky}},\ and\ \bibinfo {author} {\bibfnamefont {M.}~\bibnamefont {Olshanii}},\ }\bibfield  {title} {\bibinfo {title} {Relaxation in a completely integrable many-body quantum system: An ab initio study of the dynamics of the highly excited states of 1d lattice hard-core bosons},\ }\href {https://doi.org/10.1103/PhysRevLett.98.050405} {\bibfield  {journal} {\bibinfo  {journal} {Phys. Rev. Lett.}\ }\textbf {\bibinfo {volume} {98}},\ \bibinfo {pages} {050405} (\bibinfo {year} {2007})}\BibitemShut {NoStop}%
\bibitem [{\citenamefont {Rigol}(2009)}]{rigol2009breakdown}%
  \BibitemOpen
  \bibfield  {author} {\bibinfo {author} {\bibfnamefont {M.}~\bibnamefont {Rigol}},\ }\bibfield  {title} {\bibinfo {title} {Breakdown of thermalization in finite one-dimensional systems},\ }\href {https://doi.org/10.1103/PhysRevLett.103.100403} {\bibfield  {journal} {\bibinfo  {journal} {Phys. Rev. Lett.}\ }\textbf {\bibinfo {volume} {103}},\ \bibinfo {pages} {100403} (\bibinfo {year} {2009})}\BibitemShut {NoStop}%
\bibitem [{\citenamefont {Calabrese}\ \emph {et~al.}(2011)\citenamefont {Calabrese}, \citenamefont {Essler},\ and\ \citenamefont {Fagotti}}]{calabrese2011quantum}%
  \BibitemOpen
  \bibfield  {author} {\bibinfo {author} {\bibfnamefont {P.}~\bibnamefont {Calabrese}}, \bibinfo {author} {\bibfnamefont {F.~H.~L.}\ \bibnamefont {Essler}},\ and\ \bibinfo {author} {\bibfnamefont {M.}~\bibnamefont {Fagotti}},\ }\bibfield  {title} {\bibinfo {title} {Quantum quench in the transverse-field ising chain},\ }\href {https://doi.org/10.1103/PhysRevLett.106.227203} {\bibfield  {journal} {\bibinfo  {journal} {Phys. Rev. Lett.}\ }\textbf {\bibinfo {volume} {106}},\ \bibinfo {pages} {227203} (\bibinfo {year} {2011})}\BibitemShut {NoStop}%
\bibitem [{\citenamefont {Rigol}(2016)}]{rigol2016fundamental}%
  \BibitemOpen
  \bibfield  {author} {\bibinfo {author} {\bibfnamefont {M.}~\bibnamefont {Rigol}},\ }\bibfield  {title} {\bibinfo {title} {Fundamental asymmetry in quenches between integrable and nonintegrable systems},\ }\href {https://doi.org/10.1103/PhysRevLett.116.100601} {\bibfield  {journal} {\bibinfo  {journal} {Phys. Rev. Lett.}\ }\textbf {\bibinfo {volume} {116}},\ \bibinfo {pages} {100601} (\bibinfo {year} {2016})}\BibitemShut {NoStop}%
\bibitem [{\citenamefont {Ilievski}\ \emph {et~al.}(2015)\citenamefont {Ilievski}, \citenamefont {De~Nardis}, \citenamefont {Wouters}, \citenamefont {Caux}, \citenamefont {Essler},\ and\ \citenamefont {Prosen}}]{Ilievski2015complete}%
  \BibitemOpen
  \bibfield  {author} {\bibinfo {author} {\bibfnamefont {E.}~\bibnamefont {Ilievski}}, \bibinfo {author} {\bibfnamefont {J.}~\bibnamefont {De~Nardis}}, \bibinfo {author} {\bibfnamefont {B.}~\bibnamefont {Wouters}}, \bibinfo {author} {\bibfnamefont {J.-S.}\ \bibnamefont {Caux}}, \bibinfo {author} {\bibfnamefont {F.~H.~L.}\ \bibnamefont {Essler}},\ and\ \bibinfo {author} {\bibfnamefont {T.}~\bibnamefont {Prosen}},\ }\bibfield  {title} {\bibinfo {title} {Complete generalized gibbs ensembles in an interacting theory},\ }\href {https://doi.org/10.1103/PhysRevLett.115.157201} {\bibfield  {journal} {\bibinfo  {journal} {Phys. Rev. Lett.}\ }\textbf {\bibinfo {volume} {115}},\ \bibinfo {pages} {157201} (\bibinfo {year} {2015})}\BibitemShut {NoStop}%
\bibitem [{\citenamefont {Biagetti}\ \emph {et~al.}()\citenamefont {Biagetti}, \citenamefont {Lebek}, \citenamefont {Panfil},\ and\ \citenamefont {Nardis}}]{biagetti2408generalised}%
  \BibitemOpen
  \bibfield  {author} {\bibinfo {author} {\bibfnamefont {L.}~\bibnamefont {Biagetti}}, \bibinfo {author} {\bibfnamefont {M.}~\bibnamefont {Lebek}}, \bibinfo {author} {\bibfnamefont {M.}~\bibnamefont {Panfil}},\ and\ \bibinfo {author} {\bibfnamefont {J.}~\bibnamefont {Nardis}},\ }\bibfield  {title} {\bibinfo {title} {Generalised bbgky hierarchy for near-integrable dynamics (2024)},\ }\href@noop {} {\bibinfo  {journal} {arXiv preprint arXiv:2408.00593}\ }\BibitemShut {NoStop}%
\bibitem [{\citenamefont {Defenu}(2021)}]{defenu2021metastability}%
  \BibitemOpen
\bibfield  {journal} {  }\bibfield  {author} {\bibinfo {author} {\bibfnamefont {N.}~\bibnamefont {Defenu}},\ }\bibfield  {title} {\bibinfo {title} {Metastability and discrete spectrum of long-range systems},\ }\href {https://doi.org/10.1073/pnas.2101785118} {\bibfield  {journal} {\bibinfo  {journal} {Proceedings of the National Academy of Sciences}\ }\textbf {\bibinfo {volume} {118}},\ \bibinfo {pages} {e2101785118} (\bibinfo {year} {2021})}\BibitemShut {NoStop}%
\bibitem [{\citenamefont {Giachetti}\ and\ \citenamefont {Defenu}(2023)}]{giachetti2021entanglement}%
  \BibitemOpen
  \bibfield  {author} {\bibinfo {author} {\bibfnamefont {G.}~\bibnamefont {Giachetti}}\ and\ \bibinfo {author} {\bibfnamefont {N.}~\bibnamefont {Defenu}},\ }\bibfield  {title} {\bibinfo {title} {Entanglement propagation and dynamics in non-additive quantum systems},\ }\href {https://doi.org/10.1038/s41598-023-37984-3} {\bibfield  {journal} {\bibinfo  {journal} {Scientific Reports}\ }\textbf {\bibinfo {volume} {13}},\ \bibinfo {pages} {12388} (\bibinfo {year} {2023})}\BibitemShut {NoStop}%
\bibitem [{\citenamefont {Giachetti}\ and\ \citenamefont {Defenu}(2025)}]{giachetti2025conditions}%
  \BibitemOpen
  \bibfield  {author} {\bibinfo {author} {\bibfnamefont {G.}~\bibnamefont {Giachetti}}\ and\ \bibinfo {author} {\bibfnamefont {N.}~\bibnamefont {Defenu}},\ }\bibfield  {title} {\bibinfo {title} {Conditions for quantum violent relaxation},\ }\href@noop {} {\bibfield  {journal} {\bibinfo  {journal} {Physical Review B}\ }\textbf {\bibinfo {volume} {111}},\ \bibinfo {pages} {214301} (\bibinfo {year} {2025})}\BibitemShut {NoStop}%
\bibitem [{\citenamefont {Ba\~nuls}\ \emph {et~al.}(2011)\citenamefont {Ba\~nuls}, \citenamefont {Cirac},\ and\ \citenamefont {Hastings}}]{banuls2011strong}%
  \BibitemOpen
  \bibfield  {author} {\bibinfo {author} {\bibfnamefont {M.~C.}\ \bibnamefont {Ba\~nuls}}, \bibinfo {author} {\bibfnamefont {J.~I.}\ \bibnamefont {Cirac}},\ and\ \bibinfo {author} {\bibfnamefont {M.~B.}\ \bibnamefont {Hastings}},\ }\bibfield  {title} {\bibinfo {title} {Strong and weak thermalization of infinite nonintegrable quantum systems},\ }\href {https://doi.org/10.1103/PhysRevLett.106.050405} {\bibfield  {journal} {\bibinfo  {journal} {Phys. Rev. Lett.}\ }\textbf {\bibinfo {volume} {106}},\ \bibinfo {pages} {050405} (\bibinfo {year} {2011})}\BibitemShut {NoStop}%
\bibitem [{\citenamefont {Delfino}\ and\ \citenamefont {Viti}(2017)}]{delfino2017theory}%
  \BibitemOpen
  \bibfield  {author} {\bibinfo {author} {\bibfnamefont {G.}~\bibnamefont {Delfino}}\ and\ \bibinfo {author} {\bibfnamefont {J.}~\bibnamefont {Viti}},\ }\bibfield  {title} {\bibinfo {title} {On the theory of quantum quenches in near-critical systems},\ }\href {https://doi.org/10.1088/1751-8121/aa5660} {\bibfield  {journal} {\bibinfo  {journal} {Journal of Physics A: Mathematical and Theoretical}\ }\textbf {\bibinfo {volume} {50}},\ \bibinfo {pages} {084004} (\bibinfo {year} {2017})}\BibitemShut {NoStop}%
\bibitem [{\citenamefont {Rakovszky}\ \emph {et~al.}(2016)\citenamefont {Rakovszky}, \citenamefont {Mestyán}, \citenamefont {Collura}, \citenamefont {Kormos},\ and\ \citenamefont {Takács}}]{rakovszky2016hamiltonian}%
  \BibitemOpen
  \bibfield  {author} {\bibinfo {author} {\bibfnamefont {T.}~\bibnamefont {Rakovszky}}, \bibinfo {author} {\bibfnamefont {M.}~\bibnamefont {Mestyán}}, \bibinfo {author} {\bibfnamefont {M.}~\bibnamefont {Collura}}, \bibinfo {author} {\bibfnamefont {M.}~\bibnamefont {Kormos}},\ and\ \bibinfo {author} {\bibfnamefont {G.}~\bibnamefont {Takács}},\ }\bibfield  {title} {\bibinfo {title} {Hamiltonian truncation approach to quenches in the ising field theory},\ }\href {https://doi.org/https://doi.org/10.1016/j.nuclphysb.2016.08.024} {\bibfield  {journal} {\bibinfo  {journal} {Nuclear Physics B}\ }\textbf {\bibinfo {volume} {911}},\ \bibinfo {pages} {805} (\bibinfo {year} {2016})}\BibitemShut {NoStop}%
\bibitem [{\citenamefont {{Kormos}}\ \emph {et~al.}(2017)\citenamefont {{Kormos}}, \citenamefont {{Collura}}, \citenamefont {{Tak{\'a}cs}},\ and\ \citenamefont {{Calabrese}}}]{kormos2017real}%
  \BibitemOpen
  \bibfield  {author} {\bibinfo {author} {\bibfnamefont {M.}~\bibnamefont {{Kormos}}}, \bibinfo {author} {\bibfnamefont {M.}~\bibnamefont {{Collura}}}, \bibinfo {author} {\bibfnamefont {G.}~\bibnamefont {{Tak{\'a}cs}}},\ and\ \bibinfo {author} {\bibfnamefont {P.}~\bibnamefont {{Calabrese}}},\ }\bibfield  {title} {\bibinfo {title} {{Real-time confinement following a quantum quench to a non-integrable model}},\ }\href {https://doi.org/10.1038/nphys3934} {\bibfield  {journal} {\bibinfo  {journal} {Nat. Phys.}\ }\textbf {\bibinfo {volume} {13}},\ \bibinfo {pages} {246} (\bibinfo {year} {2017})}\BibitemShut {NoStop}%
\bibitem [{\citenamefont {Collura}\ \emph {et~al.}(2018)\citenamefont {Collura}, \citenamefont {Kormos},\ and\ \citenamefont {Tak\'acs}}]{collura2018dynamical}%
  \BibitemOpen
  \bibfield  {author} {\bibinfo {author} {\bibfnamefont {M.}~\bibnamefont {Collura}}, \bibinfo {author} {\bibfnamefont {M.}~\bibnamefont {Kormos}},\ and\ \bibinfo {author} {\bibfnamefont {G.}~\bibnamefont {Tak\'acs}},\ }\bibfield  {title} {\bibinfo {title} {Dynamical manifestation of the gibbs paradox after a quantum quench},\ }\href {https://doi.org/10.1103/PhysRevA.98.053610} {\bibfield  {journal} {\bibinfo  {journal} {Phys. Rev. A}\ }\textbf {\bibinfo {volume} {98}},\ \bibinfo {pages} {053610} (\bibinfo {year} {2018})}\BibitemShut {NoStop}%
\bibitem [{\citenamefont {Robinson}\ \emph {et~al.}(2019)\citenamefont {Robinson}, \citenamefont {James},\ and\ \citenamefont {Konik}}]{robinson2019signatures}%
  \BibitemOpen
  \bibfield  {author} {\bibinfo {author} {\bibfnamefont {N.~J.}\ \bibnamefont {Robinson}}, \bibinfo {author} {\bibfnamefont {A.~J.~A.}\ \bibnamefont {James}},\ and\ \bibinfo {author} {\bibfnamefont {R.~M.}\ \bibnamefont {Konik}},\ }\bibfield  {title} {\bibinfo {title} {Signatures of rare states and thermalization in a theory with confinement},\ }\href {https://doi.org/10.1103/PhysRevB.99.195108} {\bibfield  {journal} {\bibinfo  {journal} {Phys. Rev. B}\ }\textbf {\bibinfo {volume} {99}},\ \bibinfo {pages} {195108} (\bibinfo {year} {2019})}\BibitemShut {NoStop}%
\bibitem [{\citenamefont {Castro-Alvaredo}\ \emph {et~al.}(2020)\citenamefont {Castro-Alvaredo}, \citenamefont {Lencs\'es}, \citenamefont {Sz\'ecs\'enyi},\ and\ \citenamefont {Viti}}]{castro2020entanglement}%
  \BibitemOpen
  \bibfield  {author} {\bibinfo {author} {\bibfnamefont {O.~A.}\ \bibnamefont {Castro-Alvaredo}}, \bibinfo {author} {\bibfnamefont {M.}~\bibnamefont {Lencs\'es}}, \bibinfo {author} {\bibfnamefont {I.~M.}\ \bibnamefont {Sz\'ecs\'enyi}},\ and\ \bibinfo {author} {\bibfnamefont {J.}~\bibnamefont {Viti}},\ }\bibfield  {title} {\bibinfo {title} {Entanglement oscillations near a quantum critical point},\ }\href {https://doi.org/10.1103/PhysRevLett.124.230601} {\bibfield  {journal} {\bibinfo  {journal} {Phys. Rev. Lett.}\ }\textbf {\bibinfo {volume} {124}},\ \bibinfo {pages} {230601} (\bibinfo {year} {2020})}\BibitemShut {NoStop}%
\bibitem [{\citenamefont {Scopa}\ \emph {et~al.}(2022)\citenamefont {Scopa}, \citenamefont {Calabrese},\ and\ \citenamefont {Bastianello}}]{scopa2022entanglement}%
  \BibitemOpen
  \bibfield  {author} {\bibinfo {author} {\bibfnamefont {S.}~\bibnamefont {Scopa}}, \bibinfo {author} {\bibfnamefont {P.}~\bibnamefont {Calabrese}},\ and\ \bibinfo {author} {\bibfnamefont {A.}~\bibnamefont {Bastianello}},\ }\bibfield  {title} {\bibinfo {title} {Entanglement dynamics in confining spin chains},\ }\href {https://doi.org/10.1103/PhysRevB.105.125413} {\bibfield  {journal} {\bibinfo  {journal} {Phys. Rev. B}\ }\textbf {\bibinfo {volume} {105}},\ \bibinfo {pages} {125413} (\bibinfo {year} {2022})}\BibitemShut {NoStop}%
\bibitem [{\citenamefont {Sciolla}\ and\ \citenamefont {Biroli}(2013)}]{sciolla2013quantum}%
  \BibitemOpen
  \bibfield  {author} {\bibinfo {author} {\bibfnamefont {B.}~\bibnamefont {Sciolla}}\ and\ \bibinfo {author} {\bibfnamefont {G.}~\bibnamefont {Biroli}},\ }\bibfield  {title} {\bibinfo {title} {Quantum quenches, dynamical transitions, and off-equilibrium quantum criticality},\ }\href {https://doi.org/10.1103/PhysRevB.88.201110} {\bibfield  {journal} {\bibinfo  {journal} {Phys. Rev. B}\ }\textbf {\bibinfo {volume} {88}},\ \bibinfo {pages} {201110} (\bibinfo {year} {2013})}\BibitemShut {NoStop}%
\bibitem [{\citenamefont {Weidinger}\ \emph {et~al.}(2017)\citenamefont {Weidinger}, \citenamefont {Heyl}, \citenamefont {Silva},\ and\ \citenamefont {Knap}}]{weidinger2017dynamical}%
  \BibitemOpen
  \bibfield  {author} {\bibinfo {author} {\bibfnamefont {S.~A.}\ \bibnamefont {Weidinger}}, \bibinfo {author} {\bibfnamefont {M.}~\bibnamefont {Heyl}}, \bibinfo {author} {\bibfnamefont {A.}~\bibnamefont {Silva}},\ and\ \bibinfo {author} {\bibfnamefont {M.}~\bibnamefont {Knap}},\ }\bibfield  {title} {\bibinfo {title} {Dynamical quantum phase transitions in systems with continuous symmetry breaking},\ }\href {https://doi.org/10.1103/PhysRevB.96.134313} {\bibfield  {journal} {\bibinfo  {journal} {Phys. Rev. B}\ }\textbf {\bibinfo {volume} {96}},\ \bibinfo {pages} {134313} (\bibinfo {year} {2017})}\BibitemShut {NoStop}%
\bibitem [{\citenamefont {Chandran}\ \emph {et~al.}(2013)\citenamefont {Chandran}, \citenamefont {Nanduri}, \citenamefont {Gubser},\ and\ \citenamefont {Sondhi}}]{chandran2013equilibration}%
  \BibitemOpen
  \bibfield  {author} {\bibinfo {author} {\bibfnamefont {A.}~\bibnamefont {Chandran}}, \bibinfo {author} {\bibfnamefont {A.}~\bibnamefont {Nanduri}}, \bibinfo {author} {\bibfnamefont {S.}~\bibnamefont {Gubser}},\ and\ \bibinfo {author} {\bibfnamefont {S.}~\bibnamefont {Sondhi}},\ }\bibfield  {title} {\bibinfo {title} {Equilibration and coarsening in the quantum ${O}(n)$ model at infinite $n$},\ }\href {https://doi.org/10.1103/PhysRevB.88.024306} {\bibfield  {journal} {\bibinfo  {journal} {Phys. Rev. B}\ }\textbf {\bibinfo {volume} {88}} (\bibinfo {year} {2013})}\BibitemShut {NoStop}%
\bibitem [{\citenamefont {Moshe}\ and\ \citenamefont {Zinn-Justin}(2003)}]{moshe2003quantum}%
  \BibitemOpen
  \bibfield  {author} {\bibinfo {author} {\bibfnamefont {M.}~\bibnamefont {Moshe}}\ and\ \bibinfo {author} {\bibfnamefont {J.}~\bibnamefont {Zinn-Justin}},\ }\bibfield  {title} {\bibinfo {title} {Quantum field theory in the large n limit: a review},\ }\href {https://doi.org/10.1016/s0370-1573(03)00263-1} {\bibfield  {journal} {\bibinfo  {journal} {Physics Reports}\ }\textbf {\bibinfo {volume} {385}},\ \bibinfo {pages} {69–228} (\bibinfo {year} {2003})}\BibitemShut {NoStop}%
\bibitem [{\citenamefont {Sotiriadis}\ and\ \citenamefont {Cardy}(2010)}]{sotiriadi2010quantum}%
  \BibitemOpen
  \bibfield  {author} {\bibinfo {author} {\bibfnamefont {S.}~\bibnamefont {Sotiriadis}}\ and\ \bibinfo {author} {\bibfnamefont {J.}~\bibnamefont {Cardy}},\ }\bibfield  {title} {\bibinfo {title} {Quantum quench in interacting field theory: A self-consistent approximation},\ }\href {https://doi.org/10.1103/PhysRevB.81.134305} {\bibfield  {journal} {\bibinfo  {journal} {Phys. Rev. B}\ }\textbf {\bibinfo {volume} {81}},\ \bibinfo {pages} {134305} (\bibinfo {year} {2010})}\BibitemShut {NoStop}%
\bibitem [{\citenamefont {Vojta}(1996)}]{vojta1996quantum}%
  \BibitemOpen
  \bibfield  {author} {\bibinfo {author} {\bibfnamefont {T.}~\bibnamefont {Vojta}},\ }\bibfield  {title} {\bibinfo {title} {Quantum version of a spherical model: Crossover from quantum to classical critical behavior},\ }\href {https://doi.org/10.1103/PhysRevB.53.710} {\bibfield  {journal} {\bibinfo  {journal} {Phys. Rev. B}\ }\textbf {\bibinfo {volume} {53}},\ \bibinfo {pages} {710} (\bibinfo {year} {1996})}\BibitemShut {NoStop}%
\bibitem [{\citenamefont {Sachdev}(1999)}]{sachdev1999quantum}%
  \BibitemOpen
  \bibfield  {author} {\bibinfo {author} {\bibfnamefont {S.}~\bibnamefont {Sachdev}},\ }\href@noop {} {\emph {\bibinfo {title} {Quantum phase transitions}}},\ Vol.~\bibinfo {volume} {12}\ (\bibinfo  {publisher} {IOP Publishing},\ \bibinfo {year} {1999})\BibitemShut {NoStop}%
\bibitem [{\citenamefont {Giachetti}\ \emph {et~al.}(2025)\citenamefont {Giachetti}, \citenamefont {Solfanelli},\ and\ \citenamefont {Defenu}}]{suppmatt}%
  \BibitemOpen
  \bibfield  {author} {\bibinfo {author} {\bibfnamefont {G.}~\bibnamefont {Giachetti}}, \bibinfo {author} {\bibfnamefont {A.}~\bibnamefont {Solfanelli}},\ and\ \bibinfo {author} {\bibfnamefont {N.}~\bibnamefont {Defenu}},\ }\href@noop {} {\bibinfo {title} {Supplementary material for "universality and weak-ergodicity breaking in quantum quenches"}} (\bibinfo {year} {2025}),\ \bibinfo {note} {available as supplementary material to the main article}\BibitemShut {NoStop}%
\bibitem [{Note1()}]{Note1}%
  \BibitemOpen
  \bibinfo {note} {Notice that for $\protect \mathbf {k} = 0$, the canonical commutation relations fix $\ell _0 = 1/2$.}\BibitemShut {Stop}%
\bibitem [{\citenamefont {Choodnovsky}\ and\ \citenamefont {Choodnovsky}(1978)}]{choodnovsky1978completely}%
  \BibitemOpen
  \bibfield  {author} {\bibinfo {author} {\bibfnamefont {D.}~\bibnamefont {Choodnovsky}}\ and\ \bibinfo {author} {\bibfnamefont {G.}~\bibnamefont {Choodnovsky}},\ }\bibfield  {title} {\bibinfo {title} {Completely integrable class of mechanical systems connected with korteweg-de vries and multicomponent schr{\"o}dinger equations-i},\ }\href@noop {} {\bibfield  {journal} {\bibinfo  {journal} {S{\'e}minaire sur les {\'e}quations non lin{\'e}aires (Choodnovsky)}\ ,\ \bibinfo {pages} {1}} (\bibinfo {year} {1978})}\BibitemShut {NoStop}%
\bibitem [{\citenamefont {Wojciechowski}(1985)}]{wojciechowski1985integrability}%
  \BibitemOpen
  \bibfield  {author} {\bibinfo {author} {\bibfnamefont {S.}~\bibnamefont {Wojciechowski}},\ }\bibfield  {title} {\bibinfo {title} {Integrability of one particle in a perturbed central quartic potential},\ }\href@noop {} {\bibfield  {journal} {\bibinfo  {journal} {Physica Scripta}\ }\textbf {\bibinfo {volume} {31}},\ \bibinfo {pages} {433} (\bibinfo {year} {1985})}\BibitemShut {NoStop}%
\bibitem [{\citenamefont {Neumann}(1859)}]{neumann1859problemate}%
  \BibitemOpen
  \bibfield  {author} {\bibinfo {author} {\bibfnamefont {C.}~\bibnamefont {Neumann}},\ }\bibfield  {title} {\bibinfo {title} {De problemate quodam mechanico, quod ad primam integralium ultraellipticorum classem revocatur},\ }\href@noop {} {\bibfield  {journal} {\bibinfo  {journal} {Journal die reine und angewandte Mathematik (Crelle's Journal)}\ } (\bibinfo {year} {1859})}\BibitemShut {NoStop}%
\bibitem [{\citenamefont {Neumann}(1856)}]{neumann1856problemate}%
  \BibitemOpen
  \bibfield  {author} {\bibinfo {author} {\bibfnamefont {C.}~\bibnamefont {Neumann}},\ }\href@noop {} {\emph {\bibinfo {title} {De problemate quodam mechanico, quod ad primam integralium ultraellipticorum classem revocatur: Dissertatio inauguralis}}}\ (\bibinfo  {publisher} {Dalkowski},\ \bibinfo {year} {1856})\BibitemShut {NoStop}%
\bibitem [{\citenamefont {Uhlenbeck}(1982)}]{uhlenbeck2006equivariant}%
  \BibitemOpen
  \bibfield  {author} {\bibinfo {author} {\bibfnamefont {K.~K.}\ \bibnamefont {Uhlenbeck}},\ }\bibfield  {title} {\bibinfo {title} {Equivariant harmonic maps into spheres},\ }in\ \href@noop {} {\emph {\bibinfo {booktitle} {Harmonic Maps: Proceedings of the NSF-CBMS Regional Conference, Held at Tulane University, New Orleans December 15--19, 1980}}}\ (\bibinfo {organization} {Springer},\ \bibinfo {year} {1982})\ pp.\ \bibinfo {pages} {146--158}\BibitemShut {NoStop}%
\bibitem [{\citenamefont {Smacchia}\ \emph {et~al.}(2015)\citenamefont {Smacchia}, \citenamefont {Knap}, \citenamefont {Demler},\ and\ \citenamefont {Silva}}]{smacchia2015exploring}%
  \BibitemOpen
  \bibfield  {author} {\bibinfo {author} {\bibfnamefont {P.}~\bibnamefont {Smacchia}}, \bibinfo {author} {\bibfnamefont {M.}~\bibnamefont {Knap}}, \bibinfo {author} {\bibfnamefont {E.}~\bibnamefont {Demler}},\ and\ \bibinfo {author} {\bibfnamefont {A.}~\bibnamefont {Silva}},\ }\bibfield  {title} {\bibinfo {title} {Exploring dynamical phase transitions and prethermalization with quantum noise of excitations},\ }\href {https://doi.org/10.1103/PhysRevB.91.205136} {\bibfield  {journal} {\bibinfo  {journal} {Phys. Rev. B}\ }\textbf {\bibinfo {volume} {91}},\ \bibinfo {pages} {205136} (\bibinfo {year} {2015})}\BibitemShut {NoStop}%
\bibitem [{\citenamefont {Joyce}(1966)}]{joyce1966spherical}%
  \BibitemOpen
  \bibfield  {author} {\bibinfo {author} {\bibfnamefont {G.~S.}\ \bibnamefont {Joyce}},\ }\bibfield  {title} {\bibinfo {title} {Spherical model with long-range ferromagnetic interactions},\ }\href {https://doi.org/10.1103/PhysRev.146.349} {\bibfield  {journal} {\bibinfo  {journal} {Phys. Rev.}\ }\textbf {\bibinfo {volume} {146}},\ \bibinfo {pages} {349} (\bibinfo {year} {1966})}\BibitemShut {NoStop}%
\bibitem [{\citenamefont {Collado}\ \emph {et~al.}(2023)\citenamefont {Collado}, \citenamefont {Defenu},\ and\ \citenamefont {Lorenzana}}]{collado2023engineering}%
  \BibitemOpen
  \bibfield  {author} {\bibinfo {author} {\bibfnamefont {H.~P.~O.}\ \bibnamefont {Collado}}, \bibinfo {author} {\bibfnamefont {N.}~\bibnamefont {Defenu}},\ and\ \bibinfo {author} {\bibfnamefont {J.}~\bibnamefont {Lorenzana}},\ }\bibfield  {title} {\bibinfo {title} {Engineering higgs dynamics by spectral singularities},\ }\href {https://doi.org/10.1103/PhysRevResearch.5.023011} {\bibfield  {journal} {\bibinfo  {journal} {Phys. Rev. Res.}\ }\textbf {\bibinfo {volume} {5}},\ \bibinfo {pages} {023011} (\bibinfo {year} {2023})}\BibitemShut {NoStop}%
\bibitem [{\citenamefont {Maraga}\ \emph {et~al.}(2015)\citenamefont {Maraga}, \citenamefont {Chiocchetta}, \citenamefont {Mitra},\ and\ \citenamefont {Gambassi}}]{maraga2015aging}%
  \BibitemOpen
  \bibfield  {author} {\bibinfo {author} {\bibfnamefont {A.}~\bibnamefont {Maraga}}, \bibinfo {author} {\bibfnamefont {A.}~\bibnamefont {Chiocchetta}}, \bibinfo {author} {\bibfnamefont {A.}~\bibnamefont {Mitra}},\ and\ \bibinfo {author} {\bibfnamefont {A.}~\bibnamefont {Gambassi}},\ }\bibfield  {title} {\bibinfo {title} {Aging and coarsening in isolated quantum systems after a quench: Exact results for the quantum $\text{O}(n)$ model with $n$ $\ensuremath{\rightarrow}$ $\ensuremath{\infty}$},\ }\href {https://doi.org/10.1103/PhysRevE.92.042151} {\bibfield  {journal} {\bibinfo  {journal} {Phys. Rev. E}\ }\textbf {\bibinfo {volume} {92}},\ \bibinfo {pages} {042151} (\bibinfo {year} {2015})}\BibitemShut {NoStop}%
\bibitem [{\citenamefont {Chiocchetta}\ \emph {et~al.}(2015)\citenamefont {Chiocchetta}, \citenamefont {Tavora}, \citenamefont {Gambassi},\ and\ \citenamefont {Mitra}}]{chiocchetta2015shorttime}%
  \BibitemOpen
  \bibfield  {author} {\bibinfo {author} {\bibfnamefont {A.}~\bibnamefont {Chiocchetta}}, \bibinfo {author} {\bibfnamefont {M.}~\bibnamefont {Tavora}}, \bibinfo {author} {\bibfnamefont {A.}~\bibnamefont {Gambassi}},\ and\ \bibinfo {author} {\bibfnamefont {A.}~\bibnamefont {Mitra}},\ }\bibfield  {title} {\bibinfo {title} {Short-time universal scaling in an isolated quantum system after a quench},\ }\href {https://doi.org/10.1103/PhysRevB.91.220302} {\bibfield  {journal} {\bibinfo  {journal} {Phys. Rev. B}\ }\textbf {\bibinfo {volume} {91}},\ \bibinfo {pages} {220302} (\bibinfo {year} {2015})}\BibitemShut {NoStop}%
\bibitem [{\citenamefont {Delfino}(2020)}]{delfino2020persistent}%
  \BibitemOpen
  \bibfield  {author} {\bibinfo {author} {\bibfnamefont {G.}~\bibnamefont {Delfino}},\ }\bibfield  {title} {\bibinfo {title} {Persistent oscillations after quantum quenches: The inhomogeneous case},\ }\href {https://doi.org/10.1016/j.nuclphysb.2020.115002} {\bibfield  {journal} {\bibinfo  {journal} {Nuclear Physics B}\ }\textbf {\bibinfo {volume} {954}},\ \bibinfo {pages} {115002} (\bibinfo {year} {2020})}\BibitemShut {NoStop}%
\bibitem [{\citenamefont {Delfino}\ and\ \citenamefont {Sorba}(2022)}]{delfino2022persistent}%
  \BibitemOpen
  \bibfield  {author} {\bibinfo {author} {\bibfnamefont {G.}~\bibnamefont {Delfino}}\ and\ \bibinfo {author} {\bibfnamefont {M.}~\bibnamefont {Sorba}},\ }\bibfield  {title} {\bibinfo {title} {Persistent oscillations after quantum quenches in d dimensions},\ }\href {https://doi.org/https://doi.org/10.1016/j.nuclphysb.2021.115643} {\bibfield  {journal} {\bibinfo  {journal} {Nuclear Physics B}\ }\textbf {\bibinfo {volume} {974}},\ \bibinfo {pages} {115643} (\bibinfo {year} {2022})}\BibitemShut {NoStop}%
\bibitem [{\citenamefont {Sorba}\ \emph {et~al.}(2025)\citenamefont {Sorba}, \citenamefont {Defenu},\ and\ \citenamefont {Delfino}}]{sorba2025quantum}%
  \BibitemOpen
  \bibfield  {author} {\bibinfo {author} {\bibfnamefont {M.}~\bibnamefont {Sorba}}, \bibinfo {author} {\bibfnamefont {N.}~\bibnamefont {Defenu}},\ and\ \bibinfo {author} {\bibfnamefont {G.}~\bibnamefont {Delfino}},\ }\bibfield  {title} {\bibinfo {title} {Quantum quenches with long range interactions},\ }\href {https://doi.org/10.1016/j.nuclphysb.2025.117105} {\bibfield  {journal} {\bibinfo  {journal} {Nuclear Physics B}\ }\textbf {\bibinfo {volume} {1019}},\ \bibinfo {pages} {117105} (\bibinfo {year} {2025})}\BibitemShut {NoStop}%
\bibitem [{\citenamefont {Robertson}\ \emph {et~al.}(2024)\citenamefont {Robertson}, \citenamefont {Senese},\ and\ \citenamefont {Essler}}]{robertson2024decay}%
  \BibitemOpen
  \bibfield  {author} {\bibinfo {author} {\bibfnamefont {J.~H.}\ \bibnamefont {Robertson}}, \bibinfo {author} {\bibfnamefont {R.}~\bibnamefont {Senese}},\ and\ \bibinfo {author} {\bibfnamefont {F.~H.~L.}\ \bibnamefont {Essler}},\ }\bibfield  {title} {\bibinfo {title} {Decay of long-lived oscillations after quantum quenches in gapped interacting quantum systems},\ }\href {https://doi.org/10.1103/PhysRevA.109.032208} {\bibfield  {journal} {\bibinfo  {journal} {Phys. Rev. A}\ }\textbf {\bibinfo {volume} {109}},\ \bibinfo {pages} {032208} (\bibinfo {year} {2024})}\BibitemShut {NoStop}%
\bibitem [{\citenamefont {Balducci}\ \emph {et~al.}(2025)\citenamefont {Balducci}, \citenamefont {Chandran},\ and\ \citenamefont {Moessner}}]{balducci2025symmetry}%
  \BibitemOpen
  \bibfield  {author} {\bibinfo {author} {\bibfnamefont {F.}~\bibnamefont {Balducci}}, \bibinfo {author} {\bibfnamefont {A.}~\bibnamefont {Chandran}},\ and\ \bibinfo {author} {\bibfnamefont {R.}~\bibnamefont {Moessner}},\ }\href {https://arxiv.org/abs/2507.17386} {\bibinfo {title} {Symmetry re-breaking in an effective theory of quantum coarsening}} (\bibinfo {year} {2025}),\ \Eprint {https://arxiv.org/abs/2507.17386} {arXiv:2507.17386 [cond-mat.stat-mech]} \BibitemShut {NoStop}%
\bibitem [{Note2()}]{Note2}%
  \BibitemOpen
  \bibinfo {note} {\protect \url {https://github.com/andrea-solfanelli/Universality-and-weak-ergodicity-breaking-in-quantum-quenches}}\BibitemShut {NoStop}%
\bibitem [{\citenamefont {Debye}(1912)}]{debye1912zurtheorie}%
  \BibitemOpen
  \bibfield  {author} {\bibinfo {author} {\bibfnamefont {P.}~\bibnamefont {Debye}},\ }\bibfield  {title} {\bibinfo {title} {Zur theorie der spezifischen wärmen},\ }\href {https://doi.org/https://doi.org/10.1002/andp.19123441404} {\bibfield  {journal} {\bibinfo  {journal} {Annalen der Physik}\ }\textbf {\bibinfo {volume} {344}},\ \bibinfo {pages} {789} (\bibinfo {year} {1912})}\BibitemShut {NoStop}%
\bibitem [{\citenamefont {Ashcroft}\ and\ \citenamefont {Mermin}(1976)}]{ashcroft1976}%
  \BibitemOpen
  \bibfield  {author} {\bibinfo {author} {\bibfnamefont {N.~W.}\ \bibnamefont {Ashcroft}}\ and\ \bibinfo {author} {\bibfnamefont {N.~D.}\ \bibnamefont {Mermin}},\ }\href@noop {} {\emph {\bibinfo {title} {Solid state physics}}}\ (\bibinfo  {publisher} {Holt, Rinehart and Winston},\ \bibinfo {address} {New York},\ \bibinfo {year} {1976})\BibitemShut {NoStop}%
\end{thebibliography}

\end{widetext}
\end{document}